\documentclass{aa}  
\usepackage{graphicx}
\usepackage{txfonts}
\usepackage{lscape}
\usepackage{longtable}
\usepackage{tablefootnote}
\usepackage{placeins}
\usepackage{natbib}
\bibpunct{(}{)}{;}{a}{}{,} 
\usepackage{lscape}
\authorrunning{S. Kankkunen et al.}

\begin{document}

   \title{Long-term radio variability of active galactic nuclei at 37 GHz}

   \author{S. Kankkunen \inst{1,2}, M. Tornikoski \inst{1}, T. Hovatta \inst{3,1}, A. Lähteenmäki 
          \inst{1,2}
          }

   \institute{Aalto University Metsähovi Radio Observatory, Metsähovintie 114, FI-02540 Kylmälä, Finland
              \\
              \email{sofia.kankkunen@aalto.fi}
                \and
            Aalto University Department of Electronics and Nanoengineering, PO Box 15500, 00076 Aalto, Finland
                \and
            Finnish Centre for Astronomy with ESO, FINCA, University of Turku, Turku, FI-20014 Finland
            }

   \date{Received April 30, 2024; accepted November 11, 2024}

 
  \abstract
   {}
   {We present the results of analysing the long-term radio variability of active galactic nuclei at 37 GHz using data of 123 sources observed in the Aalto University Metsähovi Radio Observatory. Our aim was to constrain the characteristic timescales of the studied sources and to analyse whether up to 42 years of monitoring was enough to describe their variability behaviour.}
   {We used a periodogram to estimate the power spectral density of each source. The power spectral density is used to analyse the power content of a time series in the frequency domain, and it is a powerful tool in describing the variability of active galactic nuclei. We were interested in finding a bend frequency in the power spectrum, that is, a frequency at which the slope \(\beta\) of the spectrum changes from a non-zero value to zero. We fitted two models to the periodograms of each source, namely the bending power law and the simple power law. The bend frequency in the bending power law corresponds to a characteristic timescale. }
   {We were able to constrain a timescale for 11 out of 123 sources, with an average characteristic timescale \(x_b\) = 1300 days and an average power-law slope \(\beta\) = 2.3. The results suggest that up to 42 years of observations may not always be enough for obtaining a characteristic timescale in the radio domain. This is likely caused by a combination of both slow variability as well as sampling-induced effects. We also compared the obtained timescales to 43 GHz very long baseline interferometry images. The maximum length of time a knot was visible was often close to the obtained characteristic timescale. This suggests a connection between the characteristic timescale and the jet structure.}
   {}
   {}


   \maketitle
%

\section{Introduction}

The synchrotron emission of active galactic nuclei (AGNs) observed in the radio regime is generated in the relativistic jets of AGNs. When the jet is pointing close to the line of sight, the AGN is called a blazar. Blazars have highly beamed emission, and they exhibit variability in frequencies ranging from radio to \(\gamma\)-rays (e.g. \citeauthor{hovatta2019relativistic} 2019 and references therein).

Blazars and the causes for their variability have been studied intensely since their discovery in the 1960s. In blazars, the relativistic jet accelerates the synchrotron emitting electrons to such high frequencies that they are also visible in the optical range and X-rays (e.g. \citealt{urry1995unified}). 

Blazars can be divided into two categories: flat-spectrum radio quasars (FSRQs) and BL Lac objects (BLOs). The FSRQs have a spectral energy distribution (SED) where the peak is typically in the infrared regime (\citealt {hovatta2019relativistic} and references therein). In BLOs, the defining features are a flat radio spectrum with a featureless optical spectrum (\citealt{stickel1991complete}; \citealt{stocke1991einstein}). The SED peak is between the infrared and hard X-rays. A third radio-loud source type, a radio galaxy, differs from a blazar in that its relativistic jet does not point towards the line of sight. 

A popular theoretical model for explaining the variability in the jet, the shock model, was proposed by \citet{marscher1985models} and generalised to cover high-frequency radio variability in \citet{valtaoja1992five}. In their model, the injected plasma flows outwards in a conical jet, and changes in the flow parameters cause the formation of shocks. The superluminal knots, observed in blazar jets and believed to be caused by these shocks (e.g. \citealt{blandford1979relativistic}), have been temporally related to flares in AGN radio light curves (e.g. \citealt{turler2000modelling}; \citealt{savolainen2002connections}; \citealt{lindfors2006synchrotron}). 

The emission observed from AGNs is noise produced by a stochastic process. The power spectral density (PSD) of a time series describes how its power is divided over temporal frequencies. In the simplest noise-process case, the PSD shape is a simple power law, and its slope determines what type of noise it is. Here, the most relevant noise types are white noise with slope \(\beta\) = 0, flicker noise with \(\beta\) = 1, and red noise with \(\beta\) = 2 (e.g. \citealt{press1978flicker}; \citealt{uttley2002measuring}; \citealt{max2014time}; \citealt{ramakrishnan2015locating}). White noise is uncorrelated noise, whereas its integral, red noise, has uncorrelated increments. Flicker noise is found everywhere, from electronic components to biological signals and natural phenomena, but despite its prevalence, its generation mechanisms have not been successfully explained. 

The AGN variability is often studied by analysing timescales, which can in some cases be extracted from the PSD. The term 'timescale' is used ambiguously in the literature and may mean any of the following things: the average time between flares, average time from the start to finish of a flare, a periodicity, or a quasiperiodicity. A simple power law may be fitted to the PSD of a source, and the best-fit slope then describes how rapid or slow the variability of the source is. Assuming that the simple power law is sufficient in describing the noise process, one can use it as a model to study potential deviations from it. In this case any deviations would reveal a timescale that is assumed to be caused by a separate process, that is, a periodic or quasiperiodic signal with some periodic properties not consistent with the noise process.

In general, though, a simple power law may not be sufficient for describing the long-term variability of AGNs. Using the method of looking for deviations from a simple power law has been challenged due to difficulties in estimating the correct underlying noise process as well as due to challenges in estimating the significance of these deviations (\citealt{vaughan2016false}). A deviation from the underlying noise process is not the only timescale of interest though: Especially in studies using data from the X-ray regime, a bending or broken power law has been used to obtain a specific timescale, which can be seen as a bend frequency in the PSD of some sources (e.g. \citealt{edelson1999cutoff}; \citealt{uttley2002measuring}; \citealt{mchardy2006active}). In fact, it is widely accepted that this type of a bend should be present in all frequency domains but at varying timescales and that this bend represents a type of characteristic timescale (\citealt{uttley2002measuring}). Variability timescales in the X-rays are typically on the orders of days, and thus if denser sampling can be achieved, relatively short monitoring campaigns are sufficient in characterising the X-ray variability. On the other hand, AGN radio variability is difficult to characterise reliably because long-term radio monitoring is required. \citet{hovatta2007statistical} showed that for some sources, accurate analysis of their variability in the radio regime can require over 25 years of observations. To obtain such extremely long light curves with dense-enough sampling, dedicated long monitoring programmes are required. 

We generated the periodograms representing the PSDs of 123 sources from the Aalto University Metsähovi Radio Observatory (MRO) 37 GHz sample in order to probe for bend frequencies, which are physically motivated characteristic timescales. The used methods have been scrutinised thoroughly in order to make as accurate interpretations of the results as possible. An accompanying paper, hereafter Paper II, has been been written to clarify some of the common methodological pitfalls and to justify the decisions made for this analysis. 

In Sect. \ref{data}, we describe the used data and the sample. In Sect. \ref{Methods}, we explain the methods and assumptions we used in our analysis. We dedicate Sect. \ref{challenges} to some of the challenges that arise in this type of analysis, followed by the results in Sect. \ref{results}, and discussion in Sect. \ref{discussion}. We conclude the study in Sect. \ref{conclusions}. 

\section{Data and the sample} \label{data}

The MRO 14 m telescope observes AGNs in both 22 GHz and 37 GHz, where the latter includes observations of up to 42 years. The entire MRO sample consists of approximately 1200 sources. 

The long-term variability (1980-2005) of a large sample of MRO sources has been studied by \citet{hovatta2007statistical}, and the five-year variability of a smaller sample in connection with gamma-ray variability has been studied by \citet{ramakrishnan2015locating}. The data
in our study consist of observations extending until January 2023, thus bringing an additional 17 years of data to the long-term variability analysis by \citet{hovatta2007statistical}.

We opted to only use the 37 GHz observations from the MRO sample, as they are more consistently sampled than the 22 GHz observations, and they include the longest monitoring periods. The observational and reduction methods are thoroughly described in \citet{terasranta1998fifteen}. The detection limit of MRO is 200 mJy in ideal conditions. 

The sample for this study was chosen from the 37 GHz MRO sample with the following criteria: a minimum of ten years of observations, maximum flux density of at least 1 Jy, and a minimum of N = 100 data points. With these criteria, the number of sources to be analysed was 123. The sources, their total number of data points, as well as their categories (Flat-Spectrum Radio Quasar, BL Lac Object, Radio Galaxy) are listed in Table \ref{alls}.

\section{Methods} \label{Methods}

For this study, we tested different methods described thoroughly in Paper II. Below, we give the full description of the chosen analysis process with some references to Paper II, where further justification for these choices is given. The used methods are not new, but we have attempted to carefully analyse their caveats.

\subsection{The power spectral density} \label{Power Spectral Density}

The PSD of a signal describes its power content over the frequency domain. It can be obtained from the modulus squared of the discrete Fourier transform (DFT) defined as (\citealt{deeming1975fourier})

\begin{equation} \label{PSD}
|F_N(\nu)|^2 = (\sum_{i=1}^{N} f(t_i) cos(2\pi \nu t_i))^2 + (\sum_{i=1}^{N} f(t_i) sin(2\pi \nu t_i))^2,
\end{equation}
where \(f(t_i)\) is a time series, t is the time of the observation, N is the number of data points, and \(\nu\) is the sampled frequency.

In the case of a simple power-law PSD, a smooth light curve corresponds to a steep power law because there is little power in short intervals (temporal distances of flares). Conversely, an erratic light curve of short temporal distances between flares likely corresponds to a flatter slope. However, if there are large-amplitude flares over long intervals, the power law steepens even if there are also variations over short intervals. The steepness of the power-law slope depends on the overall ratio in power content. 

In general though, one should be careful in visually analysing light curves for their PSDs, as there are different ways of generating noise. \citet{press1978flicker} shows that it is possible to generate very different-looking time series that still have the same PSDs.  

\subsection{The probability density function} \label{PDF}

The probability density function (PDF) gives the distribution and probability of a random variable having a certain value. In our case, the PDF describes the distribution and probability of flux densities in a light curve. It is still unclear whether blazars have one typical PDF and how it depends on the observing frequency. The PDF of any AGN cannot be purely Gaussian, as the flux density is always positive. We discuss the PDF in the context of simulations further in Paper II. 

\subsection{Initial data processing} \label{data processing}

We cleaned the observations by only choosing the data points with a signal-to-noise ratio (S/N) of over four, as data points with S/N \(\leq\) 4 are considered non-detections in the MRO data. While it is uncommon, sometimes there are multiple observations of a given source during one day, often related to a high-activity state. Because we were not interested in intra-day variability in this study and the typical cadence of MRO observations would make such an analysis challenging, we only kept the first detection of the day. 

In addition, we shifted each observation to the beginning of the day to allow our simulation grid to be daily, and we then binned the resulting data points to be weekly. This change in the simulation grid did not have any effect on the results, but it reduced the required computing time significantly, as we were only required to simulate one data point per day instead of hourly data points. 

\subsubsection{Interpolation} \label{interpolation}

The periodogram introduced below in Sect. \ref{periodogram} expects evenly sampled data, which means that interpolation is required for our light curves. The commonly used generalized Lomb-Scargle periodogram \citep{zechmeister2009generalised}, an improved version of the Lomb-Scargle periodogram (\citealt{lomb1976least}; \citealt{scargle1982studies}), is a modification of the conventional periodogram and does not require evenly sampled data. However, we opted to not use it due to some of the issues underlined by \citet{vanderplas2018understanding}, which are also discussed in Paper II. 

We used linear interpolation for our light curves, as more complex ways to interpolate did not seem beneficial. While interpolation appeared to work satisfactory in our analysis, it is not completely without drawbacks. We discuss some of the potential issues further in Paper II. 

\cite{uttley2002measuring} observed their sources with different cadences, and they separated their PSD analysis into frequency ranges based on these observing schemes. They used the entire light curve length to determine the low-frequency portion of the PSD, the long-look portion of the light curve for the high frequencies, and the intensive monitoring segment to determine the intermediate frequencies. Because our sources have sometimes been observed with a less frequent observing cadence, specifically before the year 2000 (discussed further in Sect. \ref{flbias}), we investigated its potential effect on the obtained periodograms. In principle, our simulations should minimise the differences in the observing cadence, as we are applying the same sampling for our simulated light curves. However, some sources, notably 0716+714, have a long segment of significantly sparser sampling than the rest of the light curve. If the variability in this segment is not sufficiently well sampled, it is possible that the simulations cannot reproduce the sampling effects well enough.

We simulated data using the observing cadences of a selection of observed sources that were potentially affected and tested whether the simulations were able to replicate the behaviour. We wanted to understand whether the obtained fit for the entire light curve differed from using the shorter better-sampled segment for the higher frequencies below the bend frequency. Usually, there was no significant effect on the results. We also tested this on the observed light curves themselves to confirm they were not one of the rare extreme cases, where the variability was falsely interpreted. Due to the consistent results with both methods, we opted to use the entire light curve to determine the PSDs of each source for our analysis. 

\subsubsection{Direct current offset correction} \label{DC}

The direct current (DC) offset occurs, when a time series has a non-zero mean. This zero-frequency power then corresponds to the offset from the expected zero-mean amplitude and results in a spike in the periodogram near zero frequency. This is undesirable, because it can smear the possible real features of the power spectrum. We removed the DC offset (zero-frequency power) by subtracting the mean of the light curve. 

The subtraction of the mean may potentially cause some deviations in the position of the periodogram in the power axis. We discuss the implications and a possible solution in Paper II. 

\subsubsection{Red-noise leak and windowing} \label{rednoise}

Red-noise leak is a well-known issue in PSD analysis, when the slope is expected to be steep \(\beta\)  > 2. Red-noise leak causes the frequencies that are not the exact Fourier frequencies probed by the DFT to leak into the periodogram. Because we have a finite observing window of an on-going continuous process, we do not have the integer multiples of every possible frequency that the observed process contains. These incomplete frequencies are then spread out into our observed time frame in the frequency domain causing the periodogram to visually flatten in the higher frequencies.

This issue can be alleviated for example by using a windowing function, which reduces the side lobes of the convolution of the rectangular-shaped observing window. We chose the Hann window similarly to \citet{max2014time}. The potential downside is that the window function smears both ends of the light curve, and thus some low-frequency information is lost. This can sometimes be seen in the periodogram as a dip in low-frequency power, which affects the goodness of fit, albeit it should only increase the rejection confidence, not necessarily alter what the best-fit parameters are. We discuss both red-noise leak as well as methods for correcting it in Paper II, and show examples to demonstrate what causes the leak and the following flattening of the periodogram. 

\subsection{Monte Carlo simulations and the power spectral response method} \label{MC}

\subsubsection{The periodogram} \label{periodogram}

In order to find the characteristic timescales (bend frequencies) for the sources, we analysed their PSDs with the periodogram. To obtain the periodogram, a normalisation to the modulus squared of the light curve Fourier transform is applied. One commonly used normalisation \citep{uttley2002measuring} is
\begin{equation} \label{periodogrameq}
P(\nu) = \frac{2T}{\mu^2N^2}|F_N(\nu)|^2,
\end{equation}
where \(T\) is the length of the time series, \(\mu\) is the mean of the light curve, and \(N\) is the number of data points.
This normalisation is linked to the fractional root mean squared (RMS) variability \citep{uttley2002measuring}. The normalisation of the periodogram can be chosen quite freely when the power axis position is not of interest, which is the case in our analysis, where we are fitting only the shape of the periodogram. \citet{vaughan2003characterizing} describe some of the commonly used periodogram normalisations in astronomy. We discuss some of the issues with normalisation in Sect. \ref{challenges}. 

The periodogram is not a perfect descriptor of the PSD: The mean of the periodogram approaches the mean of the true power spectrum as the length of the time series increases. However, the variance of the true power spectrum is exaggerated by the periodogram and increasing the number of data points does not reduce it. Due to the consequential power fluctuations in the periodogram, binning of the periodogram frequencies is standard practice \citep{papadakis1993improved}. 

\subsubsection{The power spectral response method} \label{PSRESP}

The power spectral response method (PSRESP, \citealt{uttley2002measuring}) is a commonly used method for estimating the shape of the PSD of unevenly sampled AGN light curves. The sampling and other distortions of the original light curve are applied to the simulations to ensure that the simulated PSD contains all of the same non-physical effects as the PSD of the original observed light curve. 

We used the \citet{timmer1995generating} formulation for simulating the light curves. The algorithm randomises both the phase and the amplitude of the Fourier transform of the surrogate time series producing approximately Gaussian-distributed light curves, albeit the Gaussianity has recently been debated \citep{morris2019deviations} and appears to be related to the slope steepness \(\beta\). \citet{emmanoulopoulos2013generating} formulated an improved version of this method including the matching of the PDF, which closely follows the Iterated Amplitude-Adjusted Fast Fourier Transform (IAFFT) method \citep{schreiber2000surrogate}. However, we did not use the \cite{emmanoulopoulos2013generating} algorithm as the fitting of the PSD does not require a matching PDF. Instead, we exponentiated our simulated light curves to visually examine whether they appeared similar to the observed light curves when using the best-fit parameters. However, as mentioned above and discussed further in Paper II, this may not provide truly log-normal PDFs due to the possible non-Gaussianity of the original \citet{timmer1995generating} simulated light curves. Exponentiating the simulated light curves did not affect the obtained results.

\subsubsection{Models} \label{model}

\begin{figure}
    \centering
    \includegraphics[width=1\linewidth]{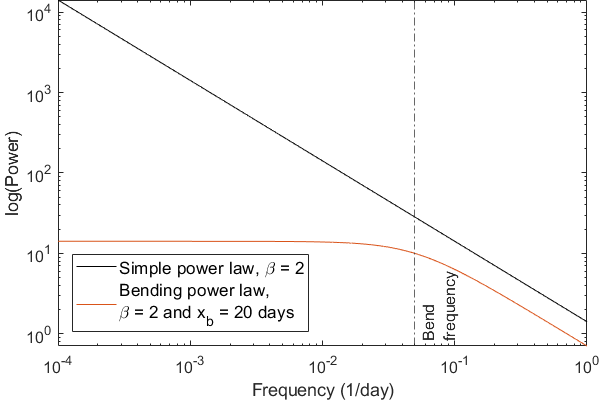}
    \caption{Comparison between the simple power law and the bending power law. The simple power law in black is a straight line of slope \(\beta_{spl}\) = 2. The bending power law in orange follows the simple power law with the same slope \(\beta\) = 2 until the bend, after which the slope turns over to \(\beta\) = 0. }
    \label{fig:models}
\end{figure}

In order to generate simulated light curves, we used both the simple power law and the bending power law. The simple power law is defined as

\begin{equation} \label{spl}
P(\nu) = \frac{1}{\nu^{\beta}},
\end{equation}
where \(\beta\) is the PSD slope.

The simple power law should provide a good fit for such sources whose characteristic timescale is long and not identifiable in the PSD. However, if the bend is already within the observing window or close to it, the simple power law may give a too flat best-fit slope. We discuss this issue further in Paper II and it is already noted in \citet{uttley2002measuring}. 

We used the following formulation for the bending power law similarly to \citealt{uttley2002measuring}: 

\begin{equation} \label{bpl}
P(\nu) = \scalebox{1.2}{$\displaystyle \frac{1}{(1+\left(\frac{\nu}{x_b^{-1}}\right)^2)^\frac{\beta}{2}}$},
\end{equation}
where \(x_b\) is the inverse of the bend frequency (characteristic timescale). The slope in this formulation flattens to zero after the bend at low frequencies. The simple power-law and bending power-law models are shown side-by-side in Fig. \ref{fig:models}. 

From previous studies, especially in the X-ray regime \citep[e.g.][and references therein]{uttley2002measuring}, we know that the light curve PSDs contain a bend. This should be the case in every frequency regime simply because we do not expect a light curve to exhibit infinite variance. Mathematically, this would mean that the behaviour is divergent, whereas convergence is required. 

We chose to use the simple power-law and bending power-law models instead of more complex models, such as one with multiple bends, as we are trying to understand whether AGNs observed in the radio domain already contain a visible bend. Cyg X-1 is an X-ray binary that has been shown to contain two bends \citep{belloni1990variability} in the hard state and thus is possibly a motivation for such behaviour in AGNs as well \citep{uttley2002measuring}. However, due to the slow variability of AGNs in the radio domain we refrain from doing such analysis now and leave it for a future study.

An important thing to note is that the bending power law is equal to the simple power law at higher frequencies. This is visible in Fig. \ref{fig:models}, where the higher frequencies converge to the same slope for both models. Because the bend is not a sharp turnover and due to sampling-induced effects, the timescale needs to be sufficiently longer than the monitoring period length for it to not have an effect on the periodogram. 

\subsubsection{Parameter space} \label{parameters}

From earlier research (e.g. \citealt{ramakrishnan2015locating}) and by visually comparing light curves between different frequency regimes, it has been established that radio variability is generally slower than in the optical, gamma-ray and X-ray frequencies. 

The typical parameters for radio variability are a slope \(\beta\) of around 2 (e.g. \citealt{max2014time}, \citealt{ramakrishnan2015locating}) and variability on timescales on the order of years \citep{hovatta2007statistical}. Because of prior information on radio variability and considering the length of each monitoring period, the searched parameter space was chosen to be \[\beta = 1:0.1:3.5 \] with a bend timescale \[x_b (days) = 500:500:7000.\] We additionally probed for higher timescales of \(x_b (days) = 100, 200, 400,\) and \(800\) to cover more of the high-frequency range.
The chosen timescales are arbitrary. We decided on this sparse slope and timescale grid due to the unevenly sampled data and to avoid the risk of overanalysing our results. The final timescale, 7000 days, is approximately half of 40 years. Of course, not all sources have 40 years of observations, but this is not an issue as there are no restrictions for fitting a bending power law with the inverse timescale (bend frequency) longer than the monitoring period. This can be understood from Fig. \ref{fig:models}, as the bending power law will simply resemble the simple power law, when moving to longer timescales.

\subsubsection{Mitigating for aliasing and red-noise leak} \label{mitig}

We simulated the light curves with a daily resolution and then binned the data points weekly exactly in the same way as the observed light curves to mimic sampling-induced deviations from the true PSD. We ensured with additional simulations that more dense sampling was not required.

Aliasing is caused by discrete sampling of a continuous process below the Nyquist frequency. Due to this insufficient sampling, power from higher frequencies leaks into the periodogram causing flattening of the high-frequency portion. Simulations with a daily resolution as well as identical binning and sampling with the observed data sufficiently mimicked this effect.

Red-noise leak causes a similar flattening effect as aliasing, but it only seriously affects larger slope values \((\beta\) > 2). We corrected for red-noise leak by simulating both 10 and 10 000 times longer light curves compared to the original observed light curves. We confirmed by tests on simulated data that 10 times longer light curves provided similar red-noise leak effects as the 10 000 times longer light curves, and we opted for the saving of computational time. 

\subsubsection{Observational errors} \label{obserr}

We drew a random Gaussian number with \(\mu\) = 0 and \(\sigma\) = 1 for each observation. We then multiplied the numbers with the standard error associated with each observation to scale the Gaussian random numbers to the correct standard deviations. These errors were then added to the simulated data to mimic observational errors.

Observational errors can be seen as a flattening of the PSD slope \(\beta\) in the highest frequencies. This limited our ability to constrain the steepest PSD slopes and it is described in more detail in Sect. \ref{results}. 

\subsection{The goodness of fit} \label{gof}

The goodness of fit is defined as \citep{uttley2002measuring}

\begin{equation}\label{chisq}
\chi_{dist}^2 = \sum_{\nu = \nu_{min}}^{\nu_{max}} \frac{(\overline{P_{sim}}(\nu)-P_{obs}(\nu))^2}{\Delta \overline{P_{sim}}(\nu)^2},
\end{equation}
where \(P_{obs}\) is the periodogram of the observed data, and \(\overline{P_{sim}}\) the mean of the simulated periodograms. The \( \chi_{dist}^2\) distribution is otherwise similar to the chi-square statistic, but it does not assume Gaussianity. 

We constructed the periodograms for both the real observations and simulations in the exact same way. We chose to use logarithmic binning with a factor of 1.3 similarly to \citet{uttley2002measuring}. We created \(\overline{P_{sim}}\) by taking the mean of all simulated light curves. 

In the PSRESP formulation, both the periodogram of the observed data as well as the periodograms of each simulated light curve are compared to the mean periodogram of the simulations. The best-fit parameters for the periodogram are found by minimising \(\chi_{dist}^2\) and calculating the percentile of the simulated \(\chi_{dist}^2\) distribution that is below the \(\chi_{dist}^2\) calculated for the observed data. This is the p-value for each parameter combination. The errors of the observed power spectrum can be estimated from the spread of the simulated periodograms around the constructed mean. The full procedure is explained in \citet{uttley2002measuring}. 

\section{Challenges} \label{challenges}

\subsection{Flare bias and uneven sampling} \label{flbias}

The MRO sample consists of unevenly sampled data. Observations are made based on source priority determined by different factors such as prior variability in the radio regime, and multi-frequency campaigns. Weather affects the observations introducing gaps and altering the telescope detection limit. Interest in high-activity sources causes the observations to be biased towards flaring events. This is challenging to apply in the Monte Carlo simulations as the PSRESP method requires replicating the sampling in order to reproduce any sampling artefacts. Flare bias may occur randomly even in the simulations, but depending on the variability, the biased sampling will also randomly occur during low-activity states. An additional issue is the sparseness of the observations before the year 2000, when a larger fraction of telescope time was used in the 22 GHz band. 

Bias towards flaring causes changes in the variance and mean, which may not be similar to the variance or mean in an evenly sampled signal. This poses some problems when simulating light curves and normalising them to match the observed light curve mean and variance. Interpolation alleviates the issue but does not completely remove it. 

A potential issue is also the artificial steepening of the periodogram slope due to interpolation. Interpolation of the light curve was shown to not substantially affect the periodogram when the light curve variability was expected to be slow and the bend timescale at an intermediate or low frequency. If the bend timescale was at a very high frequency or the variability was very fast, then the steepening of the spectrum due to the effect of interpolation became more prominent. This steepening is caused by the smoothness that interpolation causes, magnified by long gaps between observations. Flare bias complicates this issue as the effects of interpolation cannot be completely reproduced with simulations. We discuss this issue with interpolation further in Paper II.

\subsection{Periodogram normalisation} \label{pernorm}

Normalisation is also complicated by flare bias, as the variance and mean tend to be overestimated. It is also an issue due to the nature of red noise, where the parameters can vary due to statistical fluctuations (e.g. \citealt{vaughan2003characterizing}).

We attempted to normalise the periodograms to have zero mean in logarithmic space. This way, any inconsistencies in the estimate of mean or variance would not alter the position of the PSD on the power axis. However, we concluded that the steepness was often overestimated with this shifting of the periodogram and benefits over the usual normalisation were unclear. We discuss the normalisation and our results regarding it further in Paper II. 

\subsection{P-value and constraining the characteristic timescale} \label{pval}

We used a significance level of 0.1 for the bending power law and the simple power law, and we calculated the p-values using the PSRESP formulation. We report a 'rejection confidence level', that is \(1-p\), using the p-value of the best-fit parameter combination for sources for which we were able to constrain a characteristic timescale. Thus, if we have a p-value = 0.1, then the rejection confidence is 90 \%. We obtained the 90 \% confidence regions similarly to \cite{uttley2002measuring} by including all of the parameter combinations with a p-value \(p \geq\ 0.1\). While this method of obtaining the confidence region is not incorrect, it does not consider the uncertainty in both the used model and the PSD shape as the periodogram is prone to be erratic. \cite{mueller2009parameter} discuss the issues in generating confidence levels, and they suggest a different method, which is not tied to the obtained p-value. In our analysis, if the best-fit p-value is close to the chosen significance level, the confidence region may not exist, and thus it may appear as if the result is more constrained than it is.

The obtained p-value was tied to for example the factor chosen for the logarithmic binning of the periodogram, that is it varied depending on different decisions made in the data processing steps. Thus, the p-values should not be considered to be too exact. In some cases, especially with a low number of observations, many of the fits were exceptionally good. This was likely caused by interpolation as well as errors, where a low number of data points caused the light curves to be similarly smooth and errors limited the steepness of the slope. The magnitude of this effect was also dependent on the choice of both \(t_{bin}\) of the light curve and the periodogram bin size. We discuss the effect of sampling on our ability to differentiate the different PSDs further in Paper II. 

Due to Fourier frequencies, the lowest frequency bins have the lowest number of data points. This in part affects the smoothness of the periodogram around the expected characteristic timescale causing the fits to be susceptible to lower significance levels \citep{uttley2002measuring}. 

We constructed the PSDs of all sources using the periodogram, and we conducted Monte Carlo simulations to find the best-fit parameters for the PSDs of each source. We ran the simulations with 1000 iterations per parameter combination. We took a smaller sub sample and tested whether running the simulations with 10 000 iterations would alter the results, but they mostly just refined the fits. Thus, we decided to keep the n = 1000 simulations for most sources and only used the n = 10 000 iterations in the cases where a timescale appeared to be constrainable. 

We can directly compare the obtained best-fit p-values from the simple power law and the bending power law. As discussed in Sect. \ref{model}, when we move to longer bend timescales, we see that the bending power law starts to resemble the simple power law as expected. Although we fit the simple power law to the PSDs, it is equivalent to fitting a bending power law with a much longer timescale compared to the observing window. Thus, the p-values are directly comparable without using methods such as the Akaike Information Criterion to compare the models (due to a different number of parameters). If we get a reasonable fit for the simple power law, it in principle means that we cannot set an upper limit for our timescale. 

\section{Results} \label{results}

\begin{figure*}
    \centering
    \includegraphics[scale=0.6]{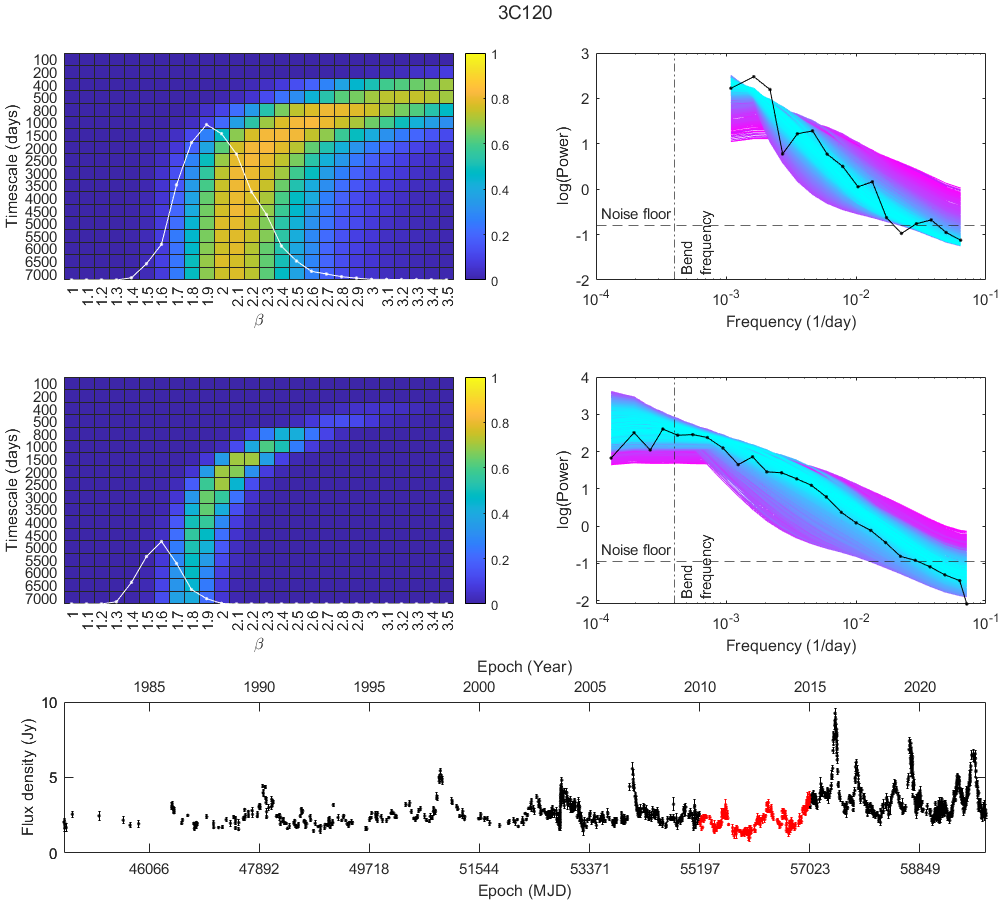}
    \caption{Comparison between the results for source 3C120, when using a shorter segment of data and the entire monitoring period. The first row of plots uses data between 2010-2015, and the middle row uses the entire light curve shown in bottom row, where the red portion corresponds to the five-year segment. The upper left-side heat map shows the PSRESP fits for each bending power-law parameter combination using five years of data. The white graph overlaid with the heat map shows the simple power-law fit with the x-axis being equivalent to the heat map x-axis. The y-axis for the white graph equals the heat map legend range scaled to agree with the label positions. The right-side plot shows the source periodogram (black) and each simulated mean periodogram with cyan indicating the best fit and magenta the worst fit. The heat map and PSD graph below show the equivalent results using the entire light curve of 3C120 shown on the bottom row. The dotted vertical lines on both periodogram graphs show the best-fit bend frequency obtained by using the entire light curve, and the dashed horizontal lines show the noise floor.}
    \label{fig:3C120}
\end{figure*}

Our results suggest that in the radio regime, many of the bend frequencies are either at such low frequencies that they are not well identifiable, or that the number of data points is not enough for an accurate PSD estimate. This does not imply that the characteristic timescale is necessarily longer than the probed length of the observing period but rather that observations must be continued for longer for the bend frequency to start dominating in the periodogram, or that more data points are required to overcome issues caused by sparse sampling. This is also apparent upon visual examination of such periodograms that are mostly consistent with a simple power law. 

Observational errors also limited our ability to constrain the slope \(\beta\). This occurs because steeper slopes correspond to smoother light curves: The white noise of the observational errors start to dominate the light curve more and more changing the shape of the periodogram back into a flatter apparent high-frequency slope. We visually examined the fits and if we saw a parameter space where the fit first worsened when going to steeper slopes and then started to improve again, we discarded those steepest \(\beta\) values. 

All of the obtained fits are listed in Table \ref{alls}. The confidence limits for the bending power-law fits are derived directly from the p-values: All fits with a p-value \(\geq\) 0.1 are considered acceptable. However, as we present them in a tabular form, we are unable to convey the fact that not all parameter combinations are possible. This can be seen below in Sect. \ref{example} and Fig. \ref{fig:3C120}, where the heat map shows how the parameters are dependent on each other. In addition, the heat maps do not convey the uncertainties in the periodogram fits themselves as discussed in Sect. \ref{pval}. The simple power-law confidence limits are similarly given based on the fits, where p \(\geq\) 0.1.

From Table \ref{alls}, we can see that for all sources, except for 1036+054, we were unable to find an upper limit for the 90 \% confidence region. This is a direct consequence of the simple power law being a good fit for most sources where the bending power law with the bend approaching infinity equals the simple power law as discussed in Sect. \ref{model} and demonstrated below. The upper limit of 1036+054 is the same as its best-fit timescale because the best-fit p-value is \(p = 0.1\). This is caused by the confidence regions being determined from the p-values. We were able to constrain a lower limit for the bend timescale for most sources.

\subsection{Interpretation of results} \label{interpres}

The characteristic timescale we are interested in is the bend frequency in the PSD. This timescale should be intrinsic to the source noise process, that is, it is not a separate periodic/quasiperiodic component, and the timescale should be related to the duration of the flares in the time series. The assumption of a relationship between flare duration and the timescale follows from our understanding of the noise process: The PSD bend frequency shows a bend after which a flare should not affect any future flares. Thus, as long as a flare is ongoing, it will affect all future flares increasing the overall variance. This matter is discussed further in Paper II and the reader is referred to for example \citet{press1978flicker} and \citet{halford1968general} for more accurate descriptions of power-law noise and its generation. 

The slope of a PSD shows the distribution of power over temporal distances in the time domain, that is, in principle, the distance between flares and what their amplitude relations are. If we have a large amount of high-amplitude short-timescale variability, the PSD slope should be flatter. If the variability appears smooth, that is the variations between long temporal distances are clearly dominant, then the PSD slope should be steeper. However, the characteristic timescale can make visual analysis more challenging as this power-law relationship continues only until a characteristic timescale, and given long-enough monitoring periods, the low-frequency portion in the PSD will become dominant. This means that given long enough monitoring periods in comparison to the characteristic timescale, all data will resemble white noise. 

For our results, we generated heat maps which show the goodness of fit of all fitted parameter combinations. The x-axis shows the slope values, and the y-axis the timescales in days. We used the heat map for our results because we found it to be the best way to convey the large number of good fits. If we only showed confidence intervals attached to singular best-fit values for the \(\beta\) and \(x_b\), we would be unable to convey how the limits in both parameters are related to each other.

\subsection{Example source 3C120} \label{example}

We chose source 3C120 as an example to demonstrate some of the concepts described earlier regarding the behaviours of the two models. This was the only source for which we analysed a shorter segment of five years in addition to the long-term analysis. We chose the shorter segment quasi-randomly, that is the segment was randomly chosen from the better-sampled portion of the light curve. 

Fig. \ref{fig:3C120} shows the results for fitting the periodogram of source 3C120 with both five years of data (2010-2015) and with using the entire light curve. The heat maps show all of the fitted parameter combinations with colour indicating the goodness of fit between 0 and 1, that is the obtained p-value. The white graph overlaid with the heat map shows the simple power-law fit with the x-axis being equivalent to the heat map x-axis. The y-axis for the simple power-law fit approximately equals the heat-map legend labels with p-value from 0 to 1, that is the higher the white graph reaches, the better the fit. 

We should get a good fit with the bending power law even if the bend is not yet visible in the PSD of the source. The issue is constraining the characteristic timescale: As can be seen in the upper left-side heat map in Fig. \ref{fig:3C120}, for five years of data, we obtain a similar best fit as for the full 40 years of data below it. However, the parameter space of near-equal fits is much larger. The periodogram of a source with a long characteristic timescale compared to the observing window length will resemble a simple power law and naturally so will the periodograms of the simulated light curves, thus causing the observed large parameter space of good fits. The p-value is 0.75 for the simple power-law fit, that is the rejection confidence is only 25 \%.

The simple power-law fit for the five years of data and 40 years of data are significantly different in both their slope-values \(\beta\) and their best-fit p-values. With five years of data, the fit is steeper (\(\beta_{spl}\) = 1.9) and closer to the value obtained by the bending power law. The best fit of the simple power-law slope is not exactly where the best-fit slopes of the bending power law are: This may occur even if the characteristic timescale is outside the observing window due to for example sampling effects or shifting and scaling of the light curve. What this means is that the bend may need to be at a significantly lower frequency than what the inverse of the observing window length is for the bend to not have any effect on the simple power-law fit. We discuss the effects of sampling and normalisation further in Paper II. The simple power law nevertheless shows a good fit also closer to the hot spot of the heat map. 

For 40 years of data, the best-fit simple-power law slope is flatter, \(\beta_{spl}\) = 1.6, consistent with our expectations of seeing a flatter fit with the periodogram, when the bend is clearly visible in the PSD. The simple power-law fit significance is also much reduced compared to the bending power-law fit: The rejection confidence for the best fit simple power-law fit \(\beta_{spl}\) = 1.6 is now 70 \%. 

These effects are also evident in both periodograms in the right-side column. There appears to be no flattening of the slope with 5 years of data, whereas with the entire light curve of over 40 years, the slope clearly flattens to a power-law slope \(\beta\) = 0. 

\subsection{Sources with well-constrained timescales and slopes}

\label{constrain}We considered the timescale and slope constrained, if the heat map of best fits showed an even distribution of values, that is there were no multiple hot spots in the heat map suggesting uncertainty in the fit. We also required for the hot spot to be sufficiently within the parameter limits so that the hot spot did not continue beyond the longest probed timescale. We also required for the simple power-law fit to be clearly worse than the bending power-law fit (minimum of 20 \% difference if both fits were poor). These were conservative decisions, and six borderline cases were left out from in-depth analysis (PKS0422+0036, OJ248, 1324+224, PKS1725+044, S52007+77, and PKS2022-077). For PKS0422+0036 the flattening of the periodogram in the low frequencies appeared to be an artefact; thus we omitted it from our analysis in this paper. 

We analysed the sources by dividing them according to their best-fit timescales. Due to the often large parameter spaces of good fits and wide bins, this is only used as an aid for comparisons. The chosen grid of timescales is not a very accurate one but sufficient for a broad estimate. For example, a timescale of 1200 days would likely have a best fit centred around \(x_b (days)\) = 1000. We chose these timescales for simplicity as they are enough to give a general idea of the source fits and should help in choosing candidates for future work with finer parameters grids. As discussed in Sect. \ref{parameters}, we also did not want to overanalyse our results because it is unclear how accurately the timescales can be determined. 

A common feature of the well-constrained sources is that their baseline flux-density levels remain fairly constant. If we assume the bend frequency to represent some function of flare duration, it would be expected for the flux density levels to return to the baseline level within the process duration which we see with these sources. A very high occurrence of flares may temporarily increase the baseline, but it should not increase infinitely as that would indicate either infinitely growing flare amplitudes or infinitely increasing occurrence of flares. These aspects are discussed in more detail in Paper II. 

Below we have divided the well-constrained sources by their bend timescales. For these 11 sources a well-constrained best fit for the bending power law was found. We provide some visual analysis of the light curves in relation to the obtained results, but wish to remind the reader that there are some caveats in this and no quantification of results should be attempted visually. 

\subsubsection{500 days}

\textit{0716+714}: Examining the light curve in Fig. \ref{fig:0716} shows short temporal distances between flares, as well as an apparent short duration of flares consistent with a short characteristic timescale. The baseline flux density does not remain completely constant, but it increases during the onset of a period with high-amplitude flares, which could be caused by their overlap.

The periodogram visually matches one with a flat slope but the bend is less evident. The bending power-law fit with a best-fit \(x_b\) = 500 days and \(\beta\) = 1.8 gives a rejection confidence of 63 \% and the simple power-law fit with \(\beta_{spl}\) = 1.2 gives a rejection confidence of 86 \%. 

\subsubsection{800 days}

\textit{2144+092}: The light curve in Fig. \ref{fig:2144} has a baseline flux density slightly above the MRO nominal detection limit with fast apparent variability but low overall flare amplitudes. This source has one of the largest parameter spaces of good fits for the constrained sources which can be, at least partially, explained by the fewer number of data points as discussed in Sect. \ref{pval}. The periodogram of the source shows a visually prominent bend and flattening into white noise at low frequencies. The bending power-law fit with a best-fit \(x_b\) = 800 days and \(\beta\) = 2.6 gives a rejection confidence of 8 \% and the simple power-law fit \(\beta_{spl}\) = 1.5 gives a rejection confidence of 66 \%.

In the heat map, the effect discussed in the beginning of this section is visible, that is the increase of fits at steeper slopes caused by observational errors. This occurs especially when the PSD slope is steep, as the observational errors are more prevalent in smoother light curves. Thus, we disregard the steepest values.

\subsubsection{1000 days}

\textit{PG0007+106}: The light curve in Fig. \ref{fig:PG0007} shows a baseline flux density close to the nominal MRO detection limit with several flares that appear to have slight overlap due to multiple peaks. The best-fit power-law slope is \(\beta\) = 2.4, and it may correspond to a light curve, where higher amplitude flares occur relatively rarely. The bending power-law fit with a best-fit \(x_b\) = 1000 days and \(\beta\) = 2.4 gives a rejection confidence of 19 \%. The simple power-law fit is rejected with over 90 \% confidence. 

Also in this source, the observational errors affect the bending power-law fit in the steeper slopes. 

\textit{0235+164}: In the heat map of Fig. \ref{fig:0235}, we see that the best-fit power-law slope is \(\beta\) = 2.4, same as the best-fit slope of the source PG0007+106. However, the light curve of this source exhibits higher amplitudes for its strongest flares and the baseline flux density is closer to 1 Jy. There is a slight dip in the low-frequency part of the periodogram, which is likely a sampling artefact or potentially caused by the Hann window. Nevertheless, visual examination of the periodogram shows an apparent bend. The bending power-law fit with a best-fit \(x_b\) = 1000 days and \(\beta\) = 2.4 gives a rejection confidence of 49 \%. The simple power-law fit is rejected with over 90 \% confidence.

\textit{0736+017}: Looking at Fig. \ref{fig:0736}, this source has similar slope values (\(\beta\) = 2.5) to the previous two sources but to the eye its light curve looks strikingly different with more erratic variations. The amplitudes of the flares are similar to PG0007+106 but its baseline flux density is higher at approximately 1 Jy. It is possible that due to the different mechanisms in which the same spectra can be generated, the flares have a smaller amplitude but a higher occurrence. Of course, it is also possible that the true timescales for the sources differ by up to 500 days or potentially more within error margins. The bending power-law fit with a best-fit \(x_b\) = 1000 days and \(\beta\) = 2.5 gives a rejection confidence of 1 \% and the simple power-law fit \(\beta_{spl}\) = 1.5 gives a rejection confidence of 75 \%. 

Visually, we see that the periodogram is very smooth explaining the low rejection confidence of the bending power-law fit. That is, with a smooth periodogram, partially obtained by chance, the fits will be high as the periodograms on average have more variations. This does not necessarily mean that the periodogram fit of this source gives a better description of the true PSD than the fits of the other sources. 

Similarly to 2144+092 and PG0007+106, the observational errors affect the fit in the steeper slope values, and we have disregarded them in our analysis. 

\textit{PKS1749+096}: The baseline flux density of the light curve varies around 2 and 3 Jy (Fig. \ref{fig:PKS1749}) and its best-fit slope is \(\beta\) = 2.1. The light curve exhibits high-amplitude flares with short temporal separations between them. Visually, the higher amplitude flares appear to occur within shorter intervals than in 0235+164 supporting the slightly flatter PSD slope. As can be seen in the heat map, this source has one of the best-constrained parameter spaces for the bending power law. The bending power-law fit with a best-fit \(x_b\) = 1000 days and \(\beta\) = 2.1 gives a rejection confidence of 10 \% and the simple power-law fit \(\beta_{spl}\) = 1.5 gives a rejection confidence of 81 \%.

\textit{2230+114}: The light curve has a baseline flux density of approximately 2 Jy and its best-fit slope is \(\beta\) = 2.3 (Fig. \ref{fig:2230}). We see multiple large flares in fairly large temporal distances from each other, supporting the steeper periodogram slope. The bending power-law fit with a best-fit \(x_b\) = 1000 days and \(\beta\) = 2.3, gives a rejection confidence of 67 \%. The simple power-law fit is rejected with over 90 \% confidence.

\subsubsection{1500 days}

\textit{0415+379}: The light curve of this source is strikingly smooth with rare high-amplitude flares, and this is also in agreement with the best-fit slope \(\beta\) = 2.8 (Fig. \ref{fig:0415}). The baseline flux density level is approximately at 2-3 Jy but appears to be decreasing slightly after the year 2010. The heat map of this source includes multiple good fits and it is not as constrained as for some of the other sources. The periodogram shows a flattening in the low frequencies preceded by a uniquely steep part in the middle of the spectrum. The bending power-law fit with a best-fit \(x_b\) = 1500 days and \(\beta\) = 2.8 gives a rejection confidence of 32 \% and the simple power law fit \(\beta_{spl}\) = 1.9 gives a rejection confidence of 69 \%.

\textit{4C29.45}: The baseline flux density of the light curve is approximately 1 Jy (Fig. \ref{fig:4C2945}). The flares have clear features and appear smooth. The slope steepness \(\beta\) = 2.5 indicates slower variability, and the light curve matches this visually. The bending power-law fit with a best-fit \(x_b\) = 1500 days and \(\beta\) = 2.5 gives a rejection confidence of 24 \% and the simple power-law fit \(\beta_{spl}\) = 1.7 gives a rejection confidence of 65 \%.

\subsubsection{2500 days}

\textit{3C120}: The baseline flux density of the light curve is approximately 2 Jy, although there is a slight increase after year 2015 (Fig. \ref{fig:3C120}). The already non-zero baseline of the flux density increases during the strongest flare and appears to remain approximately the same or slightly decreased for multiple years. The source has a best-fit slope \(\beta\) = 2.0 consistent with fairly fast variability compared to some of the other sources. The periodogram of the source shows a clear flattening in the low frequencies. The source shows almost equally good fits for timescales of 1500 days and 2000 days. The bending power-law fit with a best-fit \(x_b\) = 2500 days and \(\beta\) = 2.0 gives a rejection confidence of 29 \% and the simple power-law fit \(\beta_{spl}\) = 1.6 gives a rejection confidence of 70 \%.

\subsubsection{3000 days}

\textit{3C454.3}: The light curve is visually very striking as it has a high baseline flux density of approximately 8 Jy with moderate levels of variability, after which three prominent high-amplitude flares occur back-to-back, strongest of which exceeds 49 Jy (Fig. \ref{fig:3C4543}). After a decay period, there is another flare that is likely followed by multiple flares with high temporal frequency. The cooling down of the initial flare has visually lasted for nearly 10 years. The source has several good fits even beyond the best-fit timescale of 3000 days and the periodogram of the source shows a flattening in the lower frequencies. The bending power-law fit with with a best-fit \(x_b\) = 3000 days and \(\beta\) = 2.3 gives a rejection confidence of 21 \% and the simple power-law fit \(\beta_{spl}\) = 1.8 gives a rejection confidence of 68 \%.

\subsection{Examples of sources with unconstrained timescales}

\subsubsection{3C84}

A strikingly different source is 3C84, which already upon visual inspection of the light curve includes extremely long timescales (Fig. \ref{fig:3C84}). It is also a secondary flux calibrator in MRO due to its slow variability, especially prominent before the year 2012. 

The best simple power-law fit gives a slope \(\beta_{spl}\) = 2.6 and the best bending power-law fit a slope \(\beta\) = 3.5. This might in some cases suggest the presence of a bend due to the flattening effect caused by fitting a simple power law to a bending power law. However, as can be seen in Fig. \ref{fig:3C84}, the cause of this difference is more likely that the bend is in fact at such a long timescale that a better bending power-law fit would be found if longer timescales were included. 

The periodogram is also consistent with a simple power law. The flat region in the higher frequency part occurs when the slope itself is steep, that is the light curve is smooth. Then the observational errors dominate the region causing it to flatten. This effect is more prominent in the simple power law, where the smoothness can continue infinitely, whereas in the bending power law the timescale forces a discontinuity. 

\subsubsection{BL Lac}

For BL Lac, the best-fit bending power-law parameters are \(\beta\) = 1.9 and \(x_b\) = 7000 days (Fig. \ref{fig:BLLAC}). The simple power-law fit gives a best-fit slope \(\beta_{spl}\) = 1.7, where \(\beta_{spl}\) = 1.9 is still within error margins. For this source the slope appears fairly well constrained in the heat map but more data are required to confirm it and especially to confirm the timescale. 

\section{Discussion} \label{discussion}

As discussed briefly in Sect. \ref{results}, it is challenging to quantify an upper limit for the characteristic timescale. A simple power law will in many cases provide an adequate fit even when the bend frequency is clearly visible in the PSD of a source. This is because the high-frequency portion of the PSD equals a simple power law. Only when the power-law portion is under-represented compared to the zero-slope low-frequency portion (after the bend) should the simple power-law fit become possibly such that it can be disregarded with the usual p-value limits. In the orange graph in Fig. \ref{fig:models}, the bend is at a high frequency; thus a simple power law would clearly provide a poor fit to the periodogram. If the bend was moved to a lower frequency, there would be a larger portion of the periodogram consistent with a simple power law. This effect is magnified if no logarithmic binning is used. 

For some of the sources with a clearly better fit with the bending power law, an upper limit for the timescale may be reasonable even if it is difficult to define using the p-values. For example, the source PKS1749+096 is well sampled, and it has a clearly constrained hot spot in the heat map (Fig. \ref{fig:PKS1749}). The bending-power law fit gives a rejection confidence of only 10 \% and the simple power-law fit gives a rejection confidence of 81 \%. With the maximum fitted timescale of 7000 days, the rejection confidence is 73 \% and it is consistently increasing starting from the hot spot. Given the explanation in Sect. \ref{pval} and Sect. \ref{model}, the rejection confidence should continue increasing until such a timescale where the bending power law and simple power law converge.

Even for a source with a clear timescale, the result will always be a parameter space. The relationship between the slope \(\beta\) and the timescale \(x_b\) causes there to be a correlation between them: Given a simulated light curve of for example \(\beta\) = 2, and \(x_b\) = 500 days, the shape of the bending power law will be sufficiently similar for a good fit if we move to slightly longer timescales and a flatter slope. We discuss this further in Paper II and show examples of how sampling affects our ability to differentiate between fits. 

\subsection{Flattening to a non-zero slope}

In addition to the simple and bending power laws, \cite{uttley2002measuring} also used a high-frequency break model, where they replaced the direct flattening to \(\beta_{low}\) = 0 with a slope \(\beta_{low}\) = 1 using a sharp break. \cite{markowitz2010x} additionally used a smooth turnover model, which is more similar to the (smoothly) bending power law used in this analysis. The motivation behind using a non-zero low-frequency slope comes from studies of black hole X-ray binaries in the high state, where for example Cyg X-1 has been analysed to contain a low-frequency slope \(\beta_{low}\) = 1  (e.g. \citealt{cui1997temporal}). Here we discuss the effect of non-zero low-frequency slope on our results.

As mentioned in \cite{uttley2002measuring}, there is no physical reason for a sharp break in the PSD, and thus we decided to simulate evenly sampled light curves without observational errors using the smoothly bending power law similarly to \cite{markowitz2010x}:

\begin{equation} \label{hfbpl}
P(\nu) = \scalebox{1.2}{$\displaystyle \frac{\nu^{-\beta_{low}}}{(1+\left(\frac{\nu}{x_b^{-1}}\right))^{\beta_{high}-\beta_{low}}}$},
\end{equation}
where \(\beta_{low}\) is the low-frequency slope and \(\beta_{high}\) the high-frequency slope corresponding to \(\beta\) in Eq. \ref{bpl}.

In our simulations, we were unable to differentiate the significance of the goodness of fit between the two models irrespective of which of the two models were used as the underlying model. We used varying parameter combinations with our lowest bend timescale being 500 days. \cite{uttley2002measuring} obtained similar results using the two different break models for their observed data: Both models provided a good fit with good constraints to the p-values for their sources.

Our simulations showed a clear connection between the bend frequencies: The model with a low-frequency slope \(\beta_{low}\) = 1 would always fit a higher bend frequency to the mock light-curve PSD than if the underlying model had \(\beta_{low}\)= 0. Thus, 
using Eq. \ref{bpl} gives the most conservative (longest) bend-frequency estimate and we can be more certain that the constrained sources in fact do contain a detectable bend. If the model we use does not describe the reality well, we may leave out sources with a detectable bend in their PSDs; however, we do not obtain false positives that we consider a more serious concern. Our results also suggest that using a different low-frequency slope, \(\beta_{low}\) < 1, would affect the bend position similarly, moving it to higher frequencies. 

The high-frequency slopes obtained in our simulations were often consistent between both models, only the bend frequency position varied significantly. Thus, the slope values obtained with Eq. \ref{bpl} are likely close to the ones we would obtain using Eq. \ref{hfbpl} with a low-frequency slope \(\beta_{low}\) = 1. Due to this, we used the high-frequency slope we obtained in the simulations using Eq. \ref{bpl} to find the best-fit bend frequency when the low-frequency slope was set to \(\beta_{low}\) = 1. This should provide a reasonable approximation for the bend frequency. The results for the bend timescale with \(\beta_{low}\) = 1, using the best-fit high-frequency slope to find the bend frequency, are also reported in Table \ref{alls}.

Because we used a fixed high-frequency slope, it is expected that the obtained 90 \% confidence regions would be more constrained than when the high-frequency slope is allowed to vary freely. However, this only seems to be the case for the 11 constrained sources as well as for some of the borderline cases. This result shows how poorly we are able to constrain the timescale for most sources.

\subsection{Comparison to 43 GHz very long baseline interferometry data}

Light curves from one frequency band alone are limited in how much information they convey. As discussed further in Paper II, unless the process is Gaussian-Markov, the PSD does not reveal all statistical information of the random process \citep{press1978flicker}. In general, light curves should be compared between different frequency bands or for instance against very long baseline interferometry (VLBI) images to be able to better analyse the underlying physical conditions. 

To lay some ground work for future analysis, we compared a selection of light curves and their obtained characteristic timescales to the 43 GHz VLBI sample by Boston University (\citealt{jorstad2017kinematics}; \citealt{weaver2022kinematics})

\subsubsection{Large-scale emission}

We tested our hypothesis of how flare superposition affects the baseline flux density levels by comparing the 37 GHz emission to the 43 GHz VLBI emission of sources 3C120 and 0716+714. AGNs emit on different scales and with a single-dish telescope all emission from parsec-scale knots and kiloparsec-scale diffuse emission is visible. With VLBI, one samples only the smaller compact structures; it is not sensitive to the diffuse emission and the flux density in VLBI images is only from the parsec-scale knots. Comparing between the two light curves should allow for identification of how much the large-scale emission increases the baseline of the MRO observations. Here we assume that the large-scale emission does not exhibit any significant variability, and thus increases in the baseline flux density compared to the combined 43 GHz core+jet emission from VLBA images would then be caused by flare superposition.

\begin{figure*}
    \centering
    \includegraphics[scale=0.45]{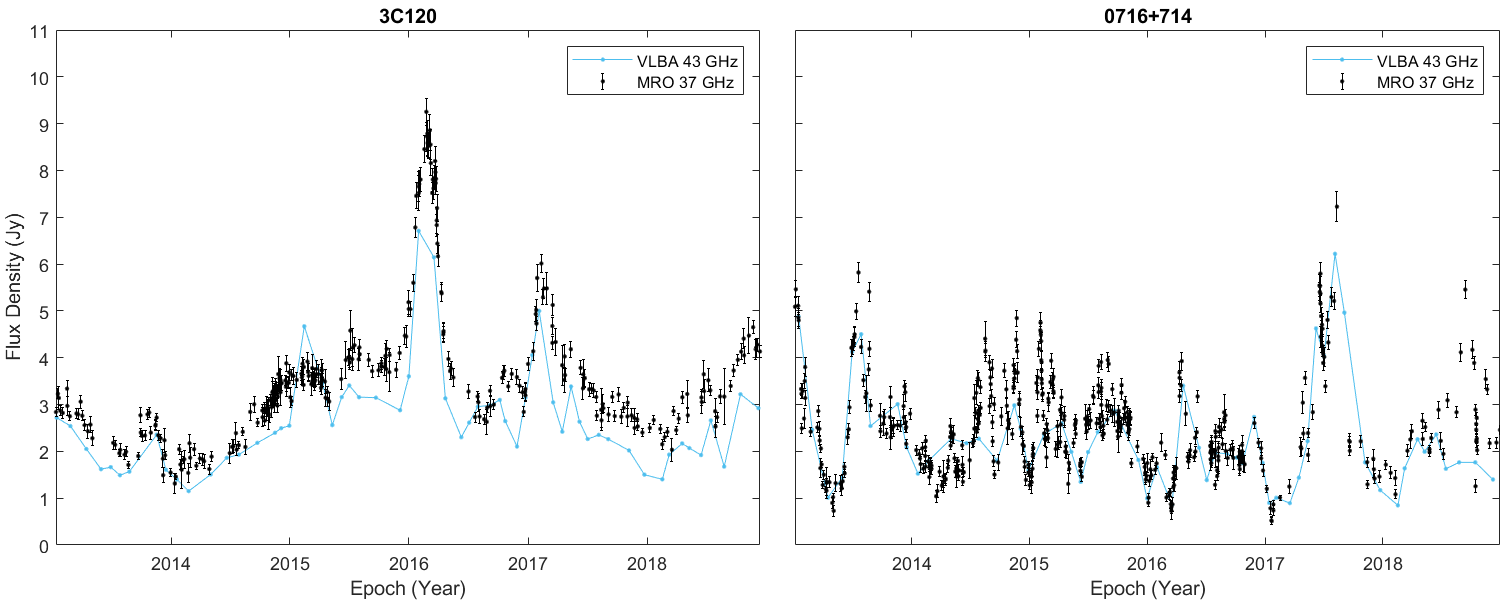}
    \caption{Light curves of 3C120 and 0716+714 using both the MRO 37 GHz data and VLBI 43 GHz (core + jet) data between 2013-2019.}
    \label{fig:vlbacomp}
\end{figure*}

We added both the core and jet flux density into a combined flux density per observation epoch and then compared it to the MRO curve in Fig. \ref{fig:vlbacomp}. The fast variability observed in 3C120 complicates the analysis as the flux density can increase over 1 Jy in one day and observations rarely coincide perfectly but rather have a few days of difference. The feature in January 2015, where the 43 GHz combined flux density exceeds 37 GHz, is likely a manifestation of this as the MRO data point is two days earlier than the VLBI 43 GHz data point. 

There are a few data points, where the flux densities were measured within 24 hours of each other. We took the average difference of these as the approximate emission excess considering that intra-day variability may cause some inaccuracies. Due to there being a limited number of such points, the error margins may be higher. 

Considering this excess, the baseline of 3C120 in 37 GHz should be at around 1 Jy, as the average difference of 10 data points was 0.92 Jy. We see it briefly decrease to that position during 2013, but otherwise it appears to remain close to 2 Jy. This may potentially be an effect caused by emission from the core and overlapping of flares.

In the case of 0716+714, the baseline drops to near the MRO nominal detection limit at multiple points despite the approximate mean-baseline level being above it (Fig. \ref{fig:0716}). One would then assume that the 43 GHz emission would be very similar to the 37 GHz light curve as it appears that there is little diffuse emission visible to the single dish observations. Indeed, both flux densities follow each other closely in the 2013-2019 segment with an average difference of 0.04 Jy for 8 data points (Fig. \ref{fig:vlbacomp}). 

\subsubsection{Knot timescales}

\begin{figure}
    \centering
    \includegraphics[width=1\linewidth]{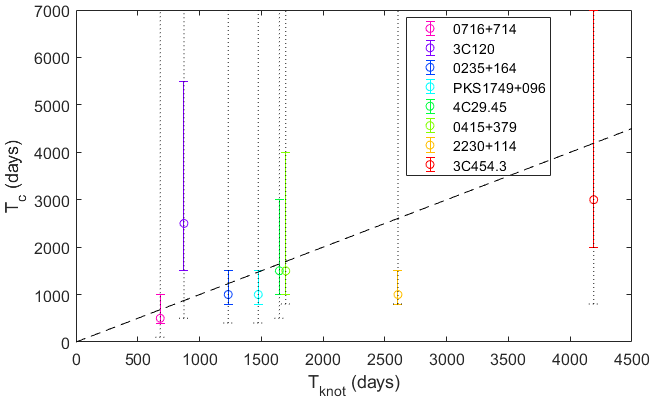}
    \caption{Scatter plot showing the results from Table \ref{tableknot}. The knot timescale is on the x-axis and the characteristic timescale on the y-axis. The coloured error bars are based on the 90 \% characteristic timescale confidence regions associated with the best-fit slope \(\beta\) obtained from the best bending power-law fit. The grey error bars are the confidence regions of the timescale when the slope is allowed to vary. For each source, except 2230+114, the knot timescale and characteristic timescale associated with the fixed best-fit slope \(\beta\) are consistent within uncertainties given the sparse parameter grid.}
    \label{fig:knotscatter}
\end{figure}

\begin{table}
\renewcommand{\arraystretch}{1.3}
\caption{Comparison between the knot timescale and characteristic timescale for applicable sources.}       
\label{tableknot}      
\centering          
\begin{tabular}{c c c c l l l }    
\hline\hline       
                     
Source & MRO alias & \(T_{knot} (days)\) & \(T_{c} (days)\)\\ 
\hline                    
   0235+164 &  & 1232    & \(1000_{800}^{1500}\)\\
   0316+413 & 3C84 & 2511 & undefined \\
   0415+379 &  & 1698 & \(1500_{1000}^{4000}\)\\  
   0430+052 & 3C120 & 874    & \(2500_{1500}^{5500}\)\\
   0716+714 &  & 682    & \(500_{400}^{1000}\)\\
   1156+295 & 4C29.45 & 1646    & \(1500_{1000}^{3000}\)\\
   1749+096 & PKS1749+096 & 1475  & \(1000_{800}^{1500}\)\\
   2200+420 & BL Lac & 4162    & undefined\\
   2230+114 &  & 2606    & \(1000_{800}^{1500}\)\\
   2251+158 & 3C454.3 & 4190    & \(3000_{2000}^{7000}\)\\
\hline           
\end{tabular}
\tablefoot{ \(T_{knot}\) gives the knot timescale from \citet{weaver2022kinematics} and \(T_{c}\) the characteristic timescale we obtained. The characteristic timescales are given with their associated 90 \% confidence regions using the fixed best-fit slope \(\beta\) obtained from the best bending power-law fit.}
\end{table}

In the radio regime, multiple authors have been able to temporally relate flaring and VLBI knots (e.g. \citealt{marscher1985models}; \citealt{turler2000modelling}; \citealt{savolainen2002connections}; \citealt{lindfors2006synchrotron}). In order to examine the connection between the obtained characteristic timescales and jet knots, we decided to find the maximum timescale an individual knot is visible from the 43 GHz sample containing approximately 10 years of data between 2008-2018. The characteristic timescale in the PSD should in principle tell us for how long a flare affects the entire light curve. Thus, it seems logical that a knot with the longest appearance would have some correlation with the bend frequency. 

We chose our sample for the comparison based on the constrained sources and our additional example sources 3C84 and BL Lac. Three sources (PG0007+106, 0736+174, and 2144+092) were not in the VLBI 43 GHz sample. We determined the approximate knot timescales by using Table 8 from \citet{weaver2022kinematics}. We compared the first observation for each knot to their last observation. The maximum knot timescale was then the maximum value found by this method. These should be considered as minimums for the longest knot timescales, that is the longest actual knot timescale may exceed these values.

Table \ref{tableknot} lists the compared sources with their characteristic timescales and VLBI knot timescales. The reported 90 \% confidence regions in Table \ref{tableknot} differ from the values in Table \ref{alls} as we are now considering only the uncertainty in the timescale associated with the best-fit slope \(\beta\) obtained from the best bending power-law fit. As discussed in Sect. \ref{results}, not all combinations of slope and timescale are possible and correspondingly the confidence regions vary. In Fig. \ref{fig:knotscatter} the timescale constraints reported in Table \ref{alls} are plotted in grey and they depict the minimum and maximum values for the timescales considering all possible slope-timescale combinations within the 90 \% confidence region. The best-fit characteristic timescales with error bars corresponding to the 90 \% confidence regions using only the obtained best-fit slope \(\beta\) are plotted in colour.

Sources 0235+164, 0415+379, 0716+714, and 4C29.45 have a best-fit timescale (near) equal to the maximum knot timescale, considering the sparse parameter grid. For all sources, except 2230+114, the timescales are close to the knot timescale within error margins. Examining the heat map of 2230+114 (Fig. \ref{fig:2230}), we see that a timescale of 2500 days - close to the knot timescale - still gives a relatively good fit using a different slope \(\beta\). However, this is the only source where the obtained timescale is not close to the knot timescale within error margins when using the best-fit slope. 

These preliminary results are encouraging and warrant further investigation into the potential connection between the characteristic timescale and the maximum time a jet knot is visible. We discuss the reasoning behind  the connection further in Paper II.

\subsection{Similar studies in the radio domain}

The 37 GHz data consists of emission arriving from closer to the central supermassive black hole compared to the two other long-term radio monitoring programmes by the University of Michigan Radio Astronomy Observatory (UMRAO, 14.5 GHz) and Owens Valley Radio Observatory (OVRO, 15 GHz). \citet{park2017long} note that for higher observing frequencies the slope of the PSD should be flatter but the characteristic timescale at a higher frequency (shorter timescale). The reason for this are the shorter cooling times of higher energy particles (e.g. \citealt{marscher1996synchrotron}).  

\subsubsection{Comparison to 15 GHz}

\citet{max2014time} used the OVRO 15 GHz data to analyse correlations between radio domain and gamma-ray data observed with the \textit{Fermi LAT Gamma-Ray Space Telescope}. They used four years of radio domain data from 2008 to 2012. \citet{park2017long} used UMRAO data (4.8, 8, and 14.5 GHz) to analyse the long-term variability of AGNs. Due to the significantly shorter light curves used by \citet{max2014time}, we only compare our results to the analysis with UMRAO data.

For some of the sources analysed by \citet{park2017long} using the lower 14.5 GHz data, the results are similar to what we obtained in our analysis. For our results, the mean of the PSD slope, when using the simple power law, was \(\overline{\beta_{spl}}\) = 1.6 which is the same result that \citeauthor{park2017long} obtained. They noted that their mean result of 1.6 was flatter than expected and that it might have to do with them using the Lomb-Scargle periodogram. However, \citet{max2014time} and \citet{ramakrishnan2015locating}, to which they compare their results to, used much shorter observing periods because they were interested in comparing the 37 GHz variability to gamma-ray variability. Thus, it is likely that the flatter mean slope is caused by the differences in the monitoring periods and that it is in fact result from the bend frequency influencing the simple power-law fit for some of the sources (see Sect. \ref{fiveyr}).

Our results are consistent with \cite{park2017long} in the number of sources that differ from the simple power law. From their smaller sample of 43 sources, only 4 were inconsistent with a simple power law. These sources are 0235+164, 3C120, 4C29.45, and 3C454.3. The sources for which we have differing results are 0415+379, 0716+714, PKS1749+096, and 2230+114. For most of these sources their best-fit simple power-law slope \(\beta\) estimate is consistent with ours varying between \(\beta\) = 1.4 and \(\beta\) = 1.6 in 14.5 GHz (or 8 GHz if the higher frequency is not available). Only for 2230+114, their best-fit slope \(\beta\) = 2.0 is clearly steeper than our simpler power-law estimate \(\beta\) = 1.6. 

\citet{park2017long} were also concerned about how well a timescale in the radio frequencies can be discovered. They simulated pulses of varying length, amplitude and arrival times and concluded that a higher degree of superposition influences the timescale, shifting it to lower frequencies. They simulated pulses with rise and decay timescales which they multiplied by a factor \(\tau\) to represent the timescale that would control the degree of overlap. Their conclusion was that the PSDs of AGN light curves would follow broken power laws intrinsically but that a high degree of superposition would move the timescale to lower frequencies.

Their analysis results are in line with what we assume the characteristic timescale to be. That is, the timescale describes how long a flare can potentially influence the light curve due to superpositioning. At some point a flare must decay sufficiently to not contribute to the light curve mean and variance anymore and that in our understanding is the characteristic timescale, which in the PSD appears as a bend frequency.

\subsubsection{Comparison to five years of MRO data} \label{fiveyr}

\citet{ramakrishnan2015locating} studied the \textit{Fermi LAT Gamma-Ray Space Telescope} \(\gamma\)-ray and 37 GHz radio correlations. They defined the simple power-law fits for 55 sources using five years of data observed by MRO. The short segment was chosen according to the coincidence with \(\gamma\)-ray observations. We would expect the five-year results to be close to the slope of the bending power law in our long-term data with constrained timescales. Indeed, the matching sources 0235+164, 0716+714, PKS1749+096, 2230+114, and 3C454.3 have the same estimated slopes (within error margins) for both five years of data and for their full monitoring period when using the bending power law, though the parameter space for the slopes is large. Interestingly though, one would expect the bend to already affect the fit especially with 0716+714, whose best-fit characteristic timescale is 500 days, well within five years of observations. The reason for this discrepancy may be in the individual decisions discussed earlier, such as bin size or scaling of light curve. 

\section{Conclusions} \label{conclusions}

We analysed the long-term radio variability of 123 sources observed by the Metsähovi Radio Observatory in 37 GHz. We utilised the longest monitoring periods in the 37 GHz band to date, maximum of which extended to 42 years of observations with an average monitoring period of 34.5 years. Our sample of 123 sources was also exceptionally large for such analysis, allowing unique insights into the long-term radio variability of AGNs.

We fitted the periodograms of each source with both the bending and simple power-law models. In addition, we compared the obtained timescales to the VLBI 43 GHz sample and the maximum knot timescales. Our main results were:

\begin{enumerate}
\item We found a well-constrained timescale and PSD slope for 11 sources. The obtained timescales varied between 500 and 3000 days and the slopes between 1.8 and 2.8.  
\item Constraining the parameters was challenging and the result was always a large parameter space. We used a sparse grid for the parameters due to the uncertainty of the benefits of using a finer grid without prior knowledge of how well a timescale can be constrained.
\item In addition to the bending power law, where the low-frequency slope flattens to zero, we used a high-frequency bend model with \(\beta_{low}\) = 1. We were unable to differentiate which model fits the data better. This appears to be a feature of the models and it may not be possible to differentiate between the two for observed radio data.
\item We found a preliminary correlation with the characteristic timescale and the maximum knot timescale of the jet. This suggests that the characteristic timescale is connected to the duration of the flares. 
\item The results from the PSD fits suggest that radio-frequency observations require long monitoring periods, sometimes exceeding 40 years of monitoring especially when the data are sparse and unevenly sampled. 
\end{enumerate}

In future work, we will analyse the constrained sources further with finer parameter grids and proper error estimates for the PSD bin powers. We will also explore the VLBI knot maximum-timescale connection with the characteristic timescale using more sources in order to confirm the found correlation statistically.

\begin{acknowledgements}
The authors would like to thank Dr. Kari Nilsson for his valuable comments on the draft. S.K. was supported by Jenny and Antti Wihuri Foundation, Väisälä Fund and Academy of Finland project 320085. T.H. was supported by Academy of Finland projects 317383, 320085, 322535, and 345899. This publication makes use of publicly available data from the Metsähovi Radio Observatory ({https://www.metsahovi.fi/opendata/}), operated by Aalto University in Finland.
\end{acknowledgements}

\bibliographystyle{aa} 
\bibliography{sources} 

\begin{appendix}
\onecolumn
\section{Heat maps and periodogram fits for the constrained sources and example comparison sources, with a result table}
\begin{figure}[h!]
    \centering
    \includegraphics[width=1\linewidth]{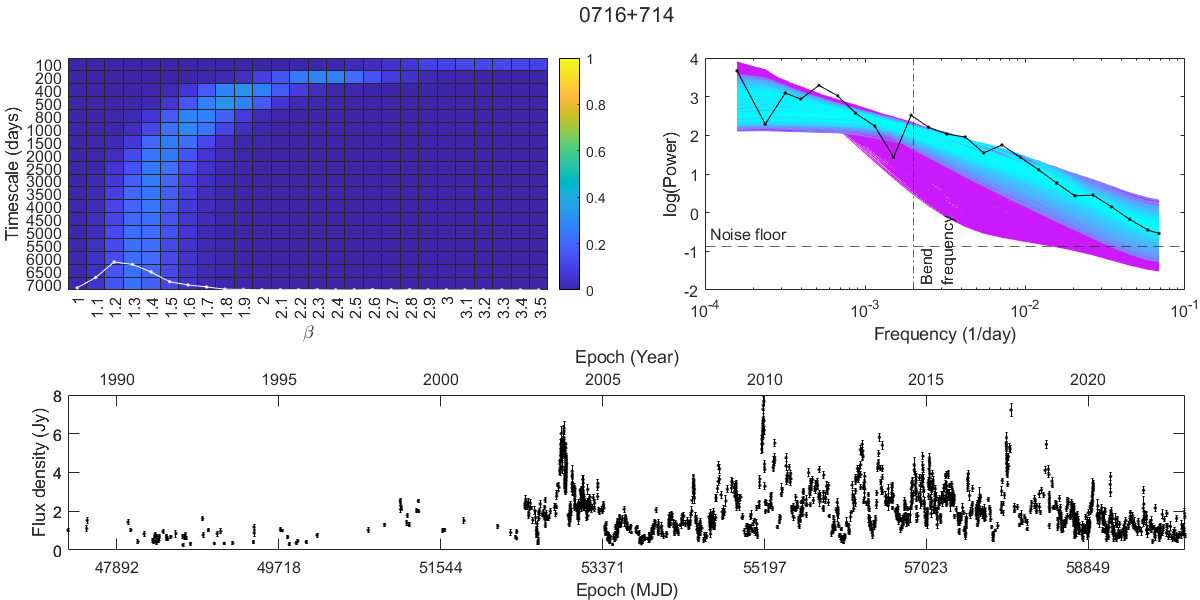}
    \caption{Results from the periodogram analysis of source 0716+714. The upper left-side plot shows the heat map overlaid with the simple power-law fit in white. The plot on its right is the periodogram of the source including all of the mean periodograms of the simulated bending power laws with cyan indicating the best fit and magenta the worst. The bottom row shows the source light curve.}
    \label{fig:0716}
\end{figure}

\begin{figure}[h!]
    \centering
    \includegraphics[width=1\linewidth]{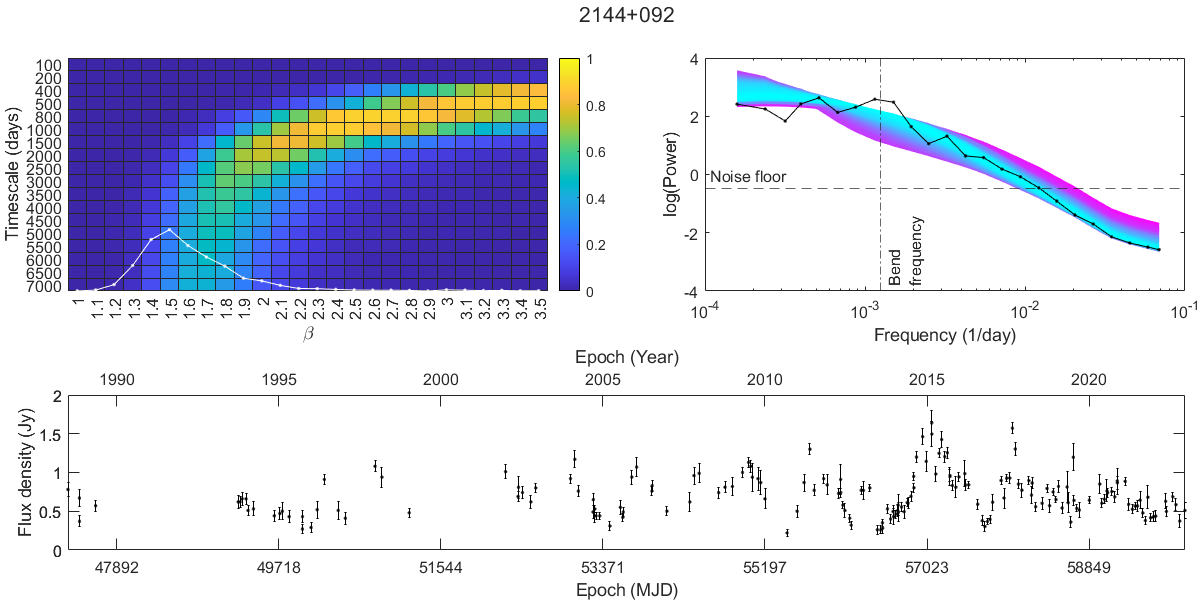}
    \caption{Results from the periodogram analysis of source 2144+092. The upper left-side plot shows the heat map  overlaid with the simple power-law fit in white. The plot on its right is the periodogram of the source including all of the mean periodograms of the simulated bending power laws with cyan indicating the best fit and magenta the worst. The bottom row shows the source light curve.}
    \label{fig:2144}
\end{figure}

\begin{figure}[h!]
    \centering
    \includegraphics[width=1\linewidth]{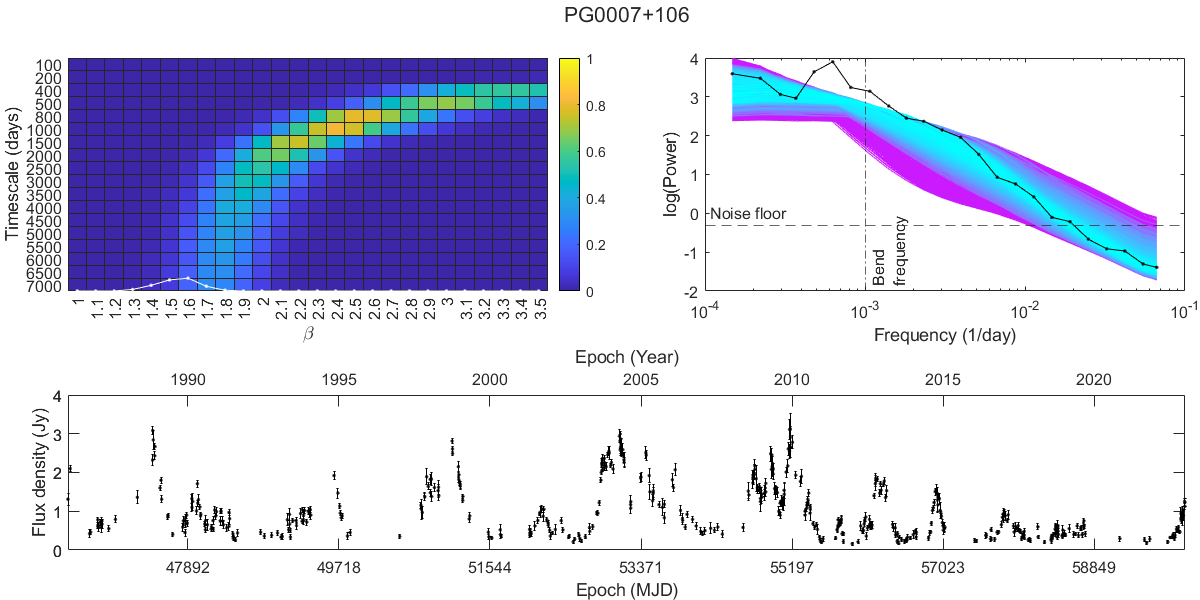}
    \caption{Results from the periodogram analysis of source PG0007+106. The upper left-side plot shows the heat map  overlaid with the simple power-law fit in white. The plot on its right is the periodogram of the source including all of the mean periodograms of the simulated bending power laws with cyan indicating the best fit and magenta the worst. The bottom row shows the source light curve.}
    \label{fig:PG0007}
\end{figure}

\begin{figure}[h!]
    \centering
    \includegraphics[width=1\linewidth]{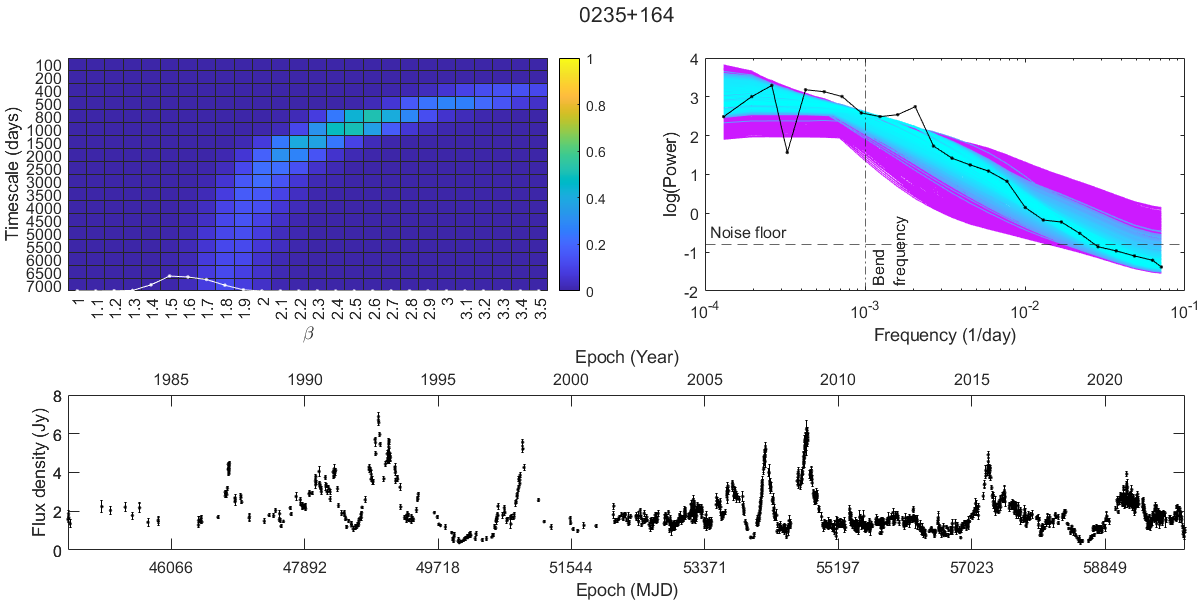}
    \caption{Results from the periodogram analysis of source 0235+164. The upper left-side plot shows the heat map  overlaid with the simple power-law fit in white. The plot on its right is the periodogram of the source including all of the mean periodograms of the simulated bending power laws with cyan indicating the best fit and magenta the worst. The bottom row shows the source light curve.}
    \label{fig:0235}
\end{figure}

\begin{figure}[h!]
    \centering
    \includegraphics[width=1\linewidth]{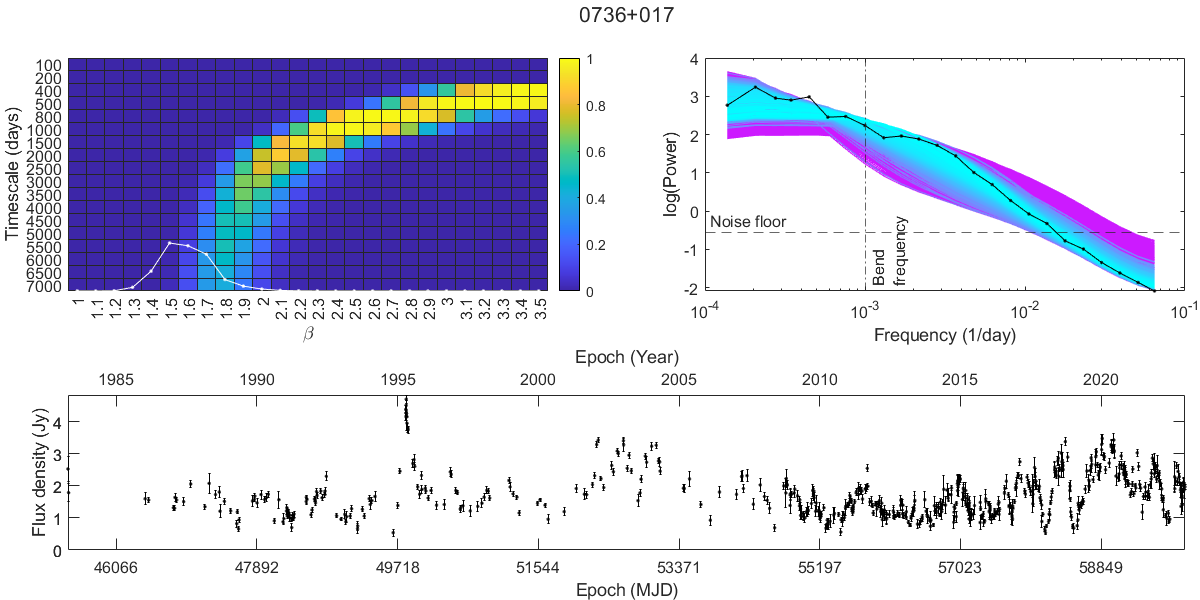}
    \caption{Results from the periodogram analysis of source 0736+017. The upper left-side plot shows the heat map overlaid with the simple power-law fit in white. The plot on its right is the periodogram of the source including all of the mean periodograms of the simulated bending power laws with cyan indicating the best fit and magenta the worst. The bottom row shows the source light curve.}
    \label{fig:0736}
\end{figure}

\begin{figure}[h!]
    \centering
    \includegraphics[width=1\linewidth]{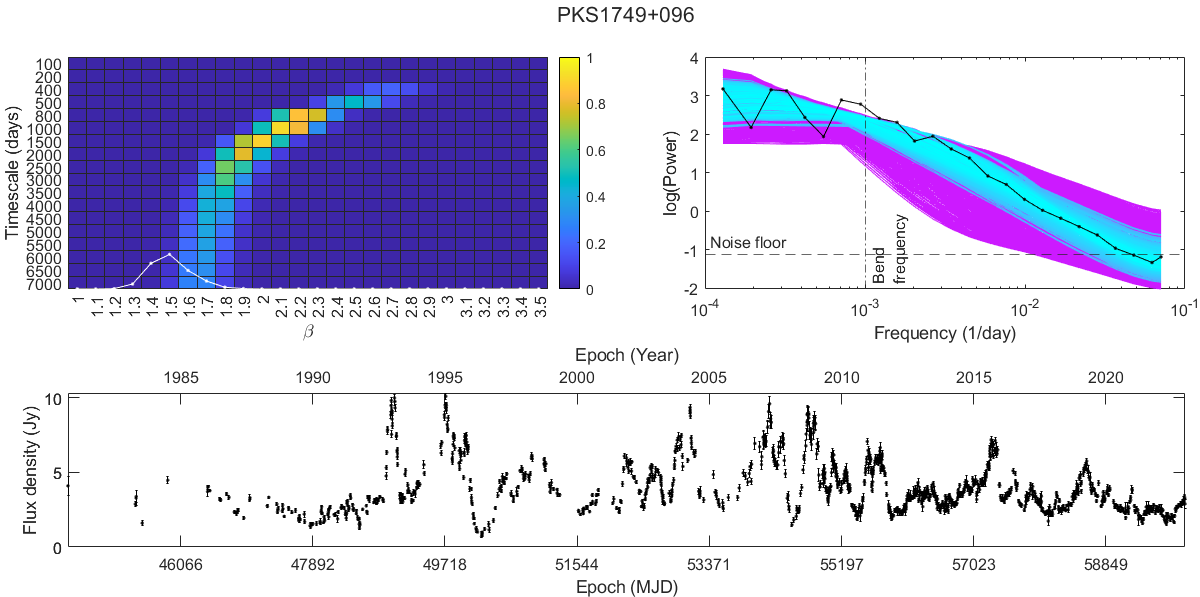}
    \caption{Results from the periodogram analysis of source PKS1749+096. The upper left-side plot shows the heat map overlaid with the simple power-law fit in white. The plot on its right is the periodogram of the source including all of the mean periodograms of the simulated bending power laws with cyan indicating the best fit and magenta the worst. The bottom row shows the source light curve.}
    \label{fig:PKS1749}
\end{figure}

\begin{figure}[h!]
    \centering
    \includegraphics[width=1\linewidth]{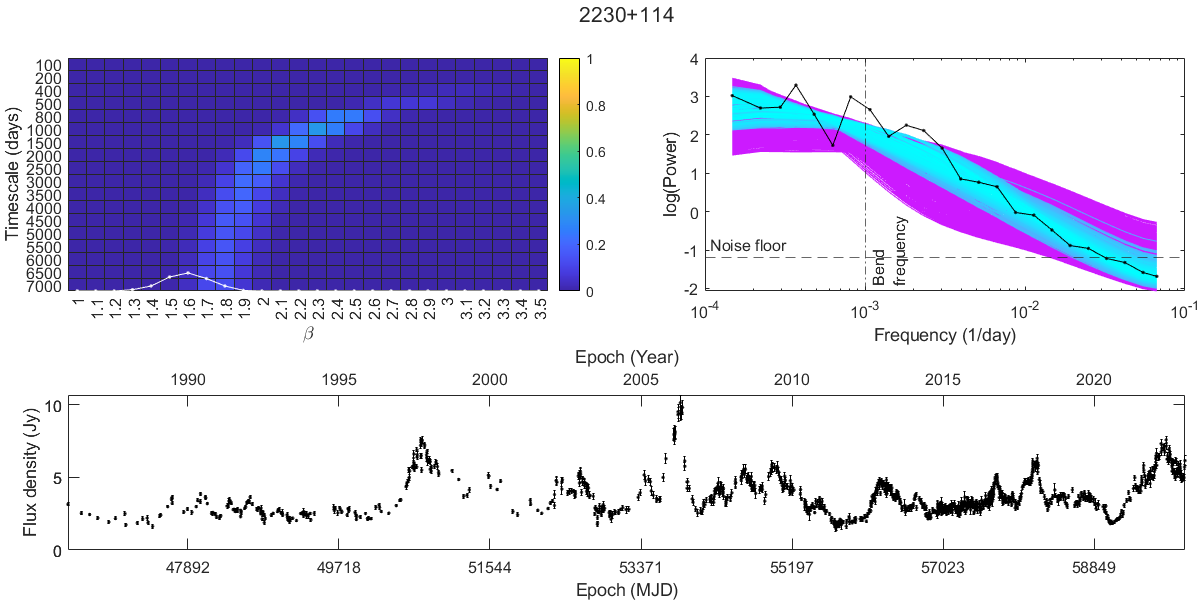}
    \caption{Results from the periodogram analysis of source 2230+114. The upper left-side plot shows the heat map overlaid with the simple power-law fit in white. The plot on its right is the periodogram of the source including all of the mean periodograms of the simulated bending power laws with cyan indicating best fit and magenta worst. The bottom row shows the source light curve.}
    \label{fig:2230}
\end{figure}

\begin{figure}[h!]
    \centering
    \includegraphics[width=1\linewidth]{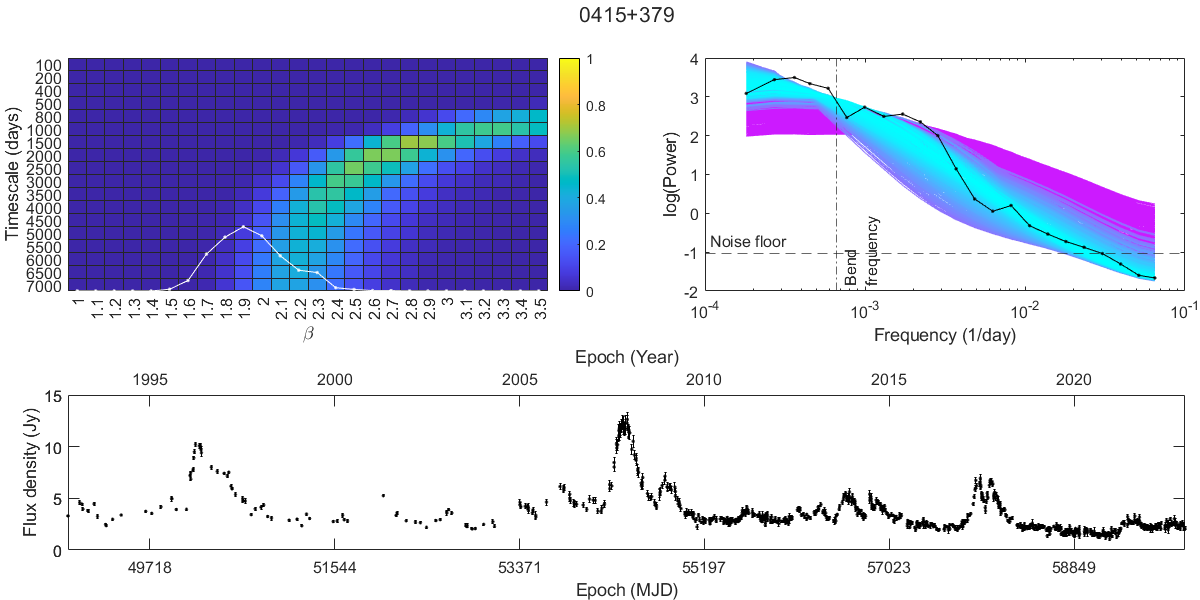}
    \caption{Results from the periodogram analysis of source 0415+379. The upper left-side plot shows the heat map overlaid with the simple power-law fit in white. The plot on its right is the periodogram of the source including all of the mean periodograms of the simulated bending power laws with cyan indicating the best fit and magenta the worst. The bottom row shows the source light curve.}
    \label{fig:0415}
\end{figure}

\begin{figure}[h!]
    \centering
    \includegraphics[width=1\linewidth]{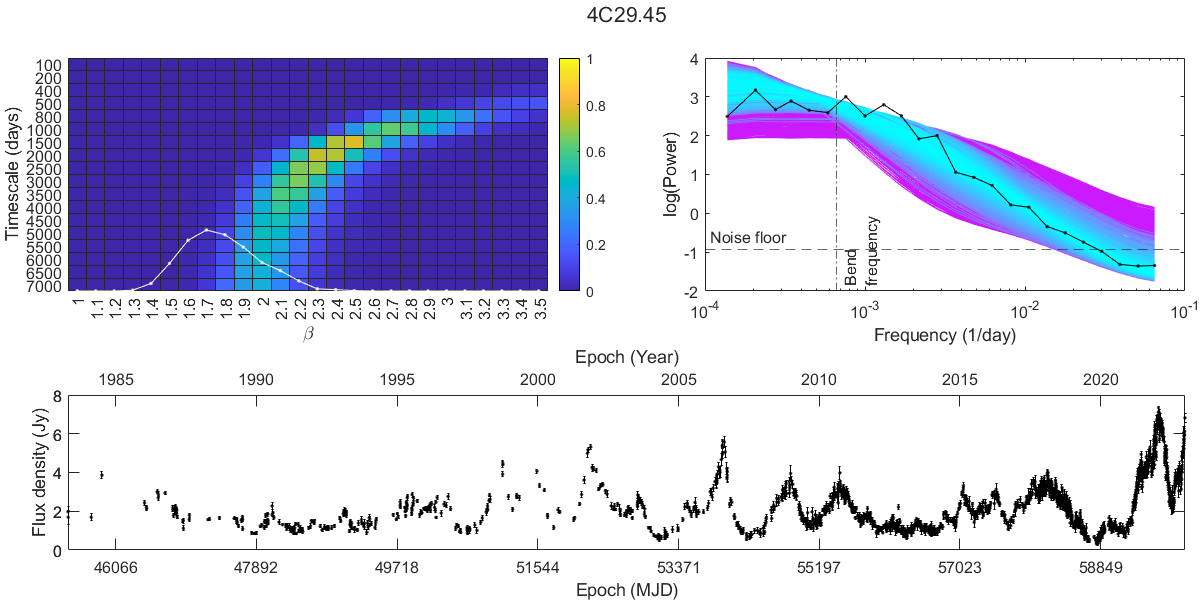}
    \caption{Results from the periodogram analysis of source 4C29.45. The upper left-side plot shows the heat map  overlaid with the simple power-law fit in white. The plot on its right is the periodogram of the source including all of the mean periodograms of the simulated bending power laws with cyan indicating the best fit and magenta the worst. The bottom row shows the source light curve.}
    \label{fig:4C2945}
\end{figure}

\begin{figure}[h!]
    \centering
    \includegraphics[width=1\linewidth]{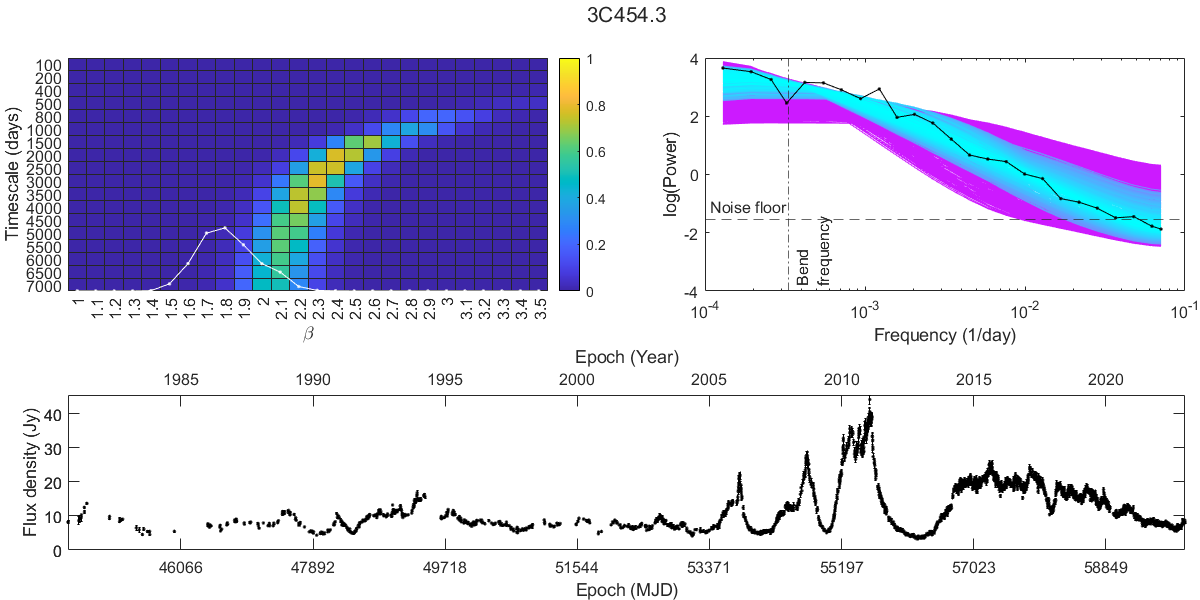}
    \caption{Results from the periodogram analysis of source 3C454.3. The upper left-side plot shows the heat map overlaid with the simple power-law fit in white. The plot on its right is the periodogram of the source including all of the mean periodograms of the simulated bending power laws with cyan indicating the best fit and magenta the worst. The bottom row shows the source light curve.}
    \label{fig:3C4543}
\end{figure}

\begin{figure}[h!]
    \centering
    \includegraphics[width=1\linewidth]{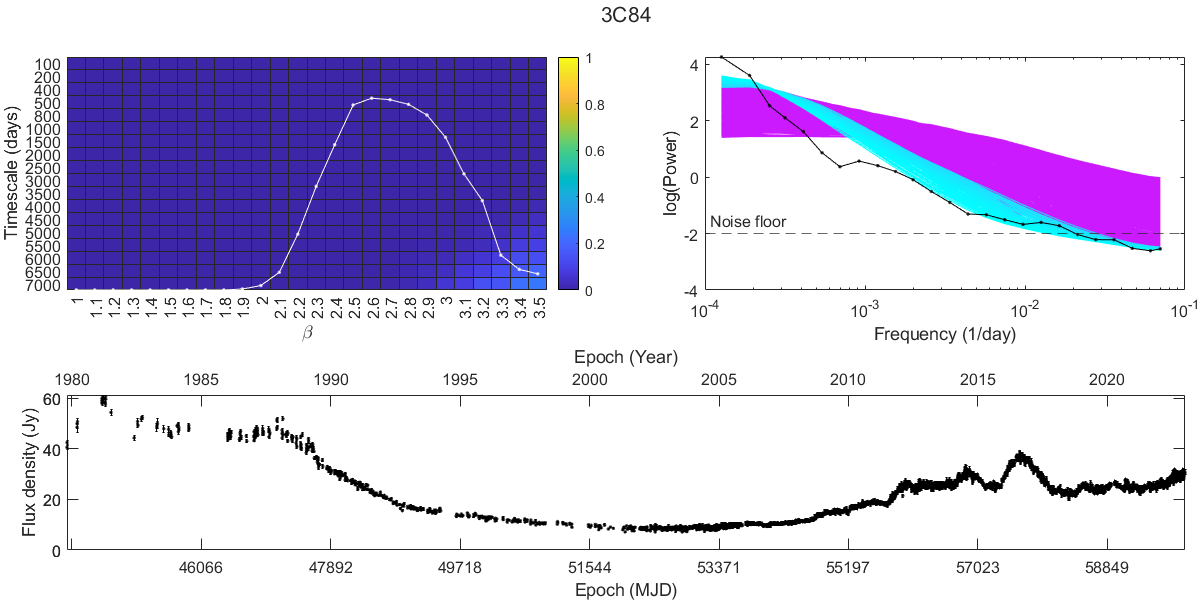}
    \caption{Results from the periodogram analysis of source 3C84. The upper left-side plot shows the heat map  overlaid with the simple power-law fit in white. The plot on its right is the periodogram of the source including all of the mean periodograms of the simulated bending power laws with cyan indicating the best fit and magenta the worst. The bottom row shows the source light curve.}
    \label{fig:3C84}
\end{figure}

\begin{figure}[h!]
    \centering
    \includegraphics[width=1\linewidth]{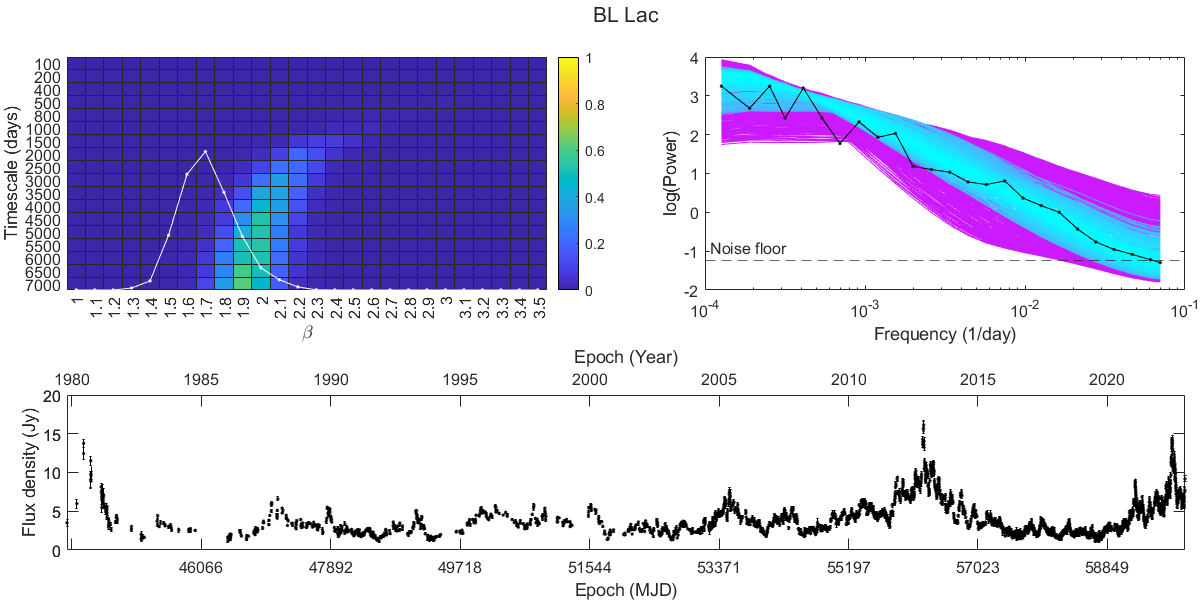}
    \caption{Results from the periodogram analysis of source BL Lac. The upper left-side plot shows the heat map overlaid with the simple power-law fit in white. The plot on its right is the periodogram of the source including all of the mean periodograms of the simulated bending power laws with cyan indicating the best fit and magenta the worst. The bottom row shows the source light curve.}
    \label{fig:BLLAC}
\end{figure}

\FloatBarrier
\renewcommand{\arraystretch}{1.3}
\begin{longtable}[h!]{llllrrrrrrr}
\caption{\label{alls} Results from the bending power-law fit (Eq. \ref{bpl}), simple power-law fit (Eq. \ref{spl}), and bending power-law fit using \(\beta_{high}\) = \(\beta_{bpl, best}\) (Eq. \ref{hfbpl}).}\\
\hline\hline
Source	&	Alias	&	N	&	Type	&	\(\beta_{bpl}\)	&	\(x_b\)	&	\(p_{bpl}\)	&	\(\beta_{spl}\)	&	\(p_{spl}\)	&	\(x_{b_1}\)	&	\(p_{bpl_1}\)	\\
\hline
\endfirsthead
\caption{Continued.}\\
\hline \hline
Source	&	Alias	&	N	&	Type	&	\(\beta_{bpl}\)	&	\(x_b\)	&	\(p_{bpl}\)	&	\(\beta_{spl}\)	&	\(p_{spl}\)	&	\(x_{b_1}\)	&	\(p_{bpl_1}\)	\\
\hline
\endhead
\hline      
\hline
0003-066	&	NRAO5	&	197	&	BLO	&	\(2.0_{1.7}^{2.6}\)	&	\(7000_{2500}^{7000}\)	&	0.44	&	\(1.8_{1.5}^{2.2}\)	&	0.69	&	\(6000_{800}^{7000}\)	&	0.51	\\
0007+106	&	PG0007+106	&	757	&	GAL	&	\(2.4_{1.6}^{3.5}\)	&	\(1000_{400}^{7000}\)	&	0.81	&		&		&	\(400_{200}^{800}\)	&	0.50	\\
0016+731	&	0016+731	&	308	&	FSRQ	&	\(1.5_{1.2}^{2.7}\)	&	\(6000_{400}^{7000}\)	&	0.78	&	\(1.4_{1.2}^{1.7}\)	&	0.53	&	\(400_{200}^{800}\)	&	0.50	\\
0048-097	&	0048-097	&	192	&	BLO	&	\(1.9_{1.4}^{3.3}\)	&	\(5500_{800}^{7000}\)	&	0.60	&	\(1.6_{1.3}^{2.2}\)	&	0.59	&	\(7000_{200}^{7000}\)	&	0.62	\\
0059+581	&	0059+581	&	629	&	FSRQ	&	\(1.7_{1.5}^{2.1}\)	&	\(6000_{1500}^{7000}\)	&	0.42	&	\(1.6_{1.4}^{1.8}\)	&	0.38	&	\(4500_{500}^{7000}\)	&	0.44	\\
0106+013	&	0106+013	&	536	&	FSRQ	&	\(2.2_{2.0}^{2.8}\)	&	\(7000_{2500}^{7000}\)	&	0.38	&	\(1.8_{1.6}^{2.5}\)	&	0.58	&	\(6500_{1500}^{7000}\)	&	0.50	\\
0109+22	&	S20109+22	&	667	&	BLO	&	\(1.8_{1.6}^{2.4}\)	&	\(6500_{1500}^{7000}\)	&	0.77	&	\(1.6_{1.3}^{2.1}\)	&	0.82	&	\(4500_{400}^{7000}\)	&	0.81	\\
0133+476	&	0133+476	&	1080	&	FSRQ	&	\(1.7_{1.5}^{2.1}\)	&	\(6000_{1500}^{7000}\)	&	0.88	&	\(1.5_{1.4}^{1.7}\)	&	0.81	&	\(6500_{800}^{7000}\)	&	0.91	\\
0149+218	&	0149+218	&	309	&	FSRQ	&	\(2.4_{1.9}^{3.5}\)	&	\(7000_{2500}^{7000}\)	&	0.27	&	\(1.8_{1.6}^{2.5}\)	&	0.43	&	\(5500_{1000}^{7000}\)	&	0.30	\\
0202+149	&	0202+149	&	399	&	FSRQ	&		&		&		&		& 	&	&	\\
0212+735	&	0212+735	&	338	&	FSRQ	&	\(1.1_{1.0}^{1.8}\)	&	\(7000_{200}^{7000}\)	&	0.66	&	\(1.1_{1.0}^{1.3}\)	&	0.74	&	\(5000_{100}^{7000}\)	&	0.76	\\
0218+357	&	0218+357	&	306	&	FSRQ	&	\(1.3_{1.2}^{1.5}\)	&	\(6500_{3500}^{7000}\)	&	0.17	&	\(1.3_{1.1}^{1.6}\)	&	0.36	&	\(6000_{100}^{7000}\)	&	0.25	\\
0219+428	&	3C66A	&	832	&	BLO	&	\(1.5_{1.3}^{1.8}\)	&	\(7000_{1500}^{7000}\)	&	0.48	&	\(1.3_{1.2}^{2.0}\)	&	0.67	&	\(4500_{100}^{7000}\)	&	0.60	\\
0224+671	&	0224+671	&	391	&	FSRQ	&	\(1.6_{1.5}^{1.8}\)	&	\(6500_{4000}^{7000}\)	&	0.19	&	\(1.5_{1.3}^{1.8}\)	&	0.58	&	\(4500_{800}^{7000}\)	&	0.27	\\
0229+131	&	0229+131	&	196	&	FSRQ	&	\(1.8_{1.3}^{3.5}\)	&	\(6500_{200}^{7000}\)	&	0.98	&	\(1.6_{1.2}^{3.1}\)	&	0.99	&	\(5000_{100}^{7000}\)	&	0.99	\\
0234+285	&	0234+285	&	914	&	FSRQ	&	\(2.1_{1.7}^{3.5}\)	&	\(5000_{800}^{7000}\)	&	0.78	&	\(1.7_{1.5}^{2.2}\)	&	0.68	&	\(3000_{800}^{7000}\)	&	0.78	\\
0235+164	&	0235+164	&	1583	&	BLO	&	\(2.4_{1.8}^{3.5}\)	&	\(1000_{400}^{7000}\)	&	0.51	&	&		&	\(400_{400}^{500}\)	&	0.26	\\
0238-084	&	0238-084	&	279	&	GAL	&	\(3.0_{1.8}^{3.5}\)	&	\(6500_{1500}^{7000}\)	&	0.89	&	\(2.2_{1.5}^{3.5}\)	&	0.89	&	\(3000_{800}^{7000}\)	&	0.91	\\
0248+430	&	0248+430	&	300	&	FSRQ	&		&		&		&		&		&	&	\\
0300+470	&	4C47.08	&	153	&	BLO	&	\(2.0_{1.6}^{3.5}\)	&	\(7000_{800}^{7000}\)	&	0.58	&	\(1.7_{1.4}^{3.1}\)	&	0.81	&	\(4500_{400}^{7000}\)	&	0.66	\\
0306+102	&	0306+102	&	238	&	FSRQ	&	\(1.6_{1.3}^{2.8}\)	&	\(6000_{400}^{7000}\)	&	0.90	&	\(1.4_{1.2}^{1.9}\)	&	0.90	&	\(6000_{100}^{7000}\)	&	0.94	\\
0316+413	&	3C84	&	9758	&	GAL	&	\(3.5_{3.2}^{3.5}\)	&	\(7000_{6000}^{7000}\)	&	0.24	&	\(2.6_{2.2}^{3.4}\)	&	0.99	&	\(7000_{3500}^{7000}\)	&	0.56	\\
0333+321	&	0333+321	&	795	&	FSRQ	&	\(2.3_{1.7}^{3.5}\)	&	\(3000_{800}^{7000}\)	&	0.78	&	\(1.7_{1.5}^{2.1}\)	&	0.50	&	\(3000_{500}^{7000}\)	&	0.68	\\
0336-019	&	CTA026	&	719	&	FSRQ	&	\(1.9_{1.6}^{3.1}\)	&	\(5500_{800}^{7000}\)	&	0.64	&	\(1.7_{1.5}^{2.1}\)	&	0.45	&	\(4000_{500}^{7000}\)	&	0.56	\\
0355+508	&	0355+508	&	1254	&	FSRQ	&	\(2.0_{1.9}^{2.3}\)	&	\(7000_{2500}^{7000}\)	&	0.36	&	\(1.7_{1.6}^{2.0}\)	&	0.47	&	\(7000_{1500}^{7000}\)	&	0.49	\\
0415+379	&	0415+379	&	891	&	GAL	&	\(2.8_{1.5}^{2.9}\)	&	\(1500_{800}^{7000}\)	&	0.68	&	\(1.9_{1.7}^{2.2}\)	&	0.31	&	\(800_{400}^{2000}\)	&	0.64	\\
0420-014	&	0420-014	&	1094	&	FSRQ	&	\(2.0_{1.9}^{2.2}\)	&	\(7000_{3500}^{7000}\)	&	0.29	&	\(1.7_{1.6}^{1.9}\)	&	0.27	&	\(6000_{2000}^{7000}\)	&	0.29	\\
0422+0036	&	PKS0422+0036	&	409	&	BLO	&	\(2.7_{1.6}^{3.5}\)	&	\(1000_{400}^{7000}\)	&	0.63	&	\(1.6_{1.5}^{1.8}\)	&	0.25	&	\(6000_{2000}^{7000}\)	&	0.29	\\
0430+052	&	3C120	&	1539	&	GAL	&	\(2.0_{1.6}^{2.9}\)	&	\(2500_{500}^{7000}\)	&	0.71	& \(1.6_{1.4}^{1.7}\)	&	0.30	&	\(1000_{400}^{4500}\)	&	0.56	\\
0440-003	&	NRAO190	&	232	&	FSRQ	&	\(1.9_{1.3}^{3.5}\)	&	\(6500_{400}^{7000}\)	&	0.95	&	\(1.6_{1.2}^{3.2}\)	&	0.94	&	\(4500_{100}^{7000}\)	&	0.95	\\
0446+112	&	PKS0446+112	&	237	&	GAL	&	\(1.7_{1.5}^{2.1}\)	&	\(6000_{2000}^{7000}\)	&	0.38	&	\(1.5_{1.3}^{1.8}\)	&	0.42	&	\(7000_{400}^{7000}\)	&	0.41	\\
0458-020	&	PKS0458-020	&	355	&	FSRQ	&	\(2.4_{2.1}^{3.5}\)	&	\(7000_{5500}^{7000}\)	&	0.17	&	\(2.0_{1.7}^{3.3}\)	&	0.73	&	\(7000_{2000}^{7000}\)	&	0.29	\\
0507+179	&	0507+179	&	175	&	FSRQ	&	\(1.8_{1.6}^{2.6}\)	&	\(7000_{1000}^{7000}\)	&	0.44	&	\(1.7_{1.5}^{1.9}\)	&	0.44	&	\(7000_{1000}^{7000}\)	&	0.45	\\
0528+134	&	0528+134	&	1031	&	FSRQ	&	\(2.4_{2.1}^{3.2}\)	&	\(7000_{2000}^{7000}\)	&	0.74	&	\(2.0_{1.8}^{2.3}\)	&	0.66	&	\(5000_{1500}^{7000}\)	&	0.79	\\
0552+398	&	0552+398	&	939	&	FSRQ	&		&	&		&	\(2.8_{2.0}^{3.5}\)	&	0.55	&	&	\\
0605-085	&	PKS0605-085	&	267	&	FSRQ	&	&	&	&	\(1.6_{1.5}^{1.7}\)	&	0.14	&		&	\\
0642+449	&	0642+449	&	841	&	FSRQ	&	\(1.7_{1.6}^{2.1}\)	&	\(7000_{2500}^{7000}\)	&	0.60	&	\(1.6_{1.4}^{2.0}\)	&	0.90	&	\(7000_{800}^{7000}\)	&	0.84	\\
0716+714	&	0716+714	&	1934	&	BLO	&	\(1.8_{1.1}^{2.6}\)	&	\(500_{200}^{7000}\)	&	0.37	&	\(1.2_{1.2}^{1.3}\)	&	0.14	&	\(100_{100}^{400}\)	&	0.32	\\
0723-008	&	PKS0723-008	&	184	&	BLO	&	\(1.7_{1.5}^{2.1}\)	&	\(7000_{1500}^{7000}\)	&	0.47	&	\(1.6_{1.4}^{1.9}\)	&	0.67	&	\(7000_{400}^{7000}\)	&	0.57	\\
0735+17	&	PKS0735+17	&	932	&	BLO	&	\(3.1_{2.2}^{3.5}\)	&	\(7000_{3000}^{7000}\)	&	0.67	&	\(2.1_{1.7}^{3.1}\)	&	0.99	&	\(7000_{1500}^{7000}\)	&	0.85	\\
0736+017	&	0736+017	&	608	&	FSRQ	&	\(2.5_{1.6}^{3.5}\)	&	\(1000_{400}^{7000}\)	&	0.99	&	\(1.5_{1.4}^{1.7}\)	&	0.25	&	\(200_{100}^{500}\)	&	0.99	\\
0748+126	&	0748+126	&	187	&	FSRQ	&	\(2.0_{1.6}^{3.5}\)	&	\(6500_{1500}^{7000}\)	&	0.64	&	\(1.8_{1.5}^{2.8}\)	&	0.85	&	\(7000_{800}^{7000}\)	&	0.66	\\
0754+100	&	0754+100	&	379	&	BLO	&	\(1.5_{1.4}^{2.4}\)	&	\(5500_{500}^{7000}\)	&	0.40	&	\(1.4_{1.3}^{1.4}\)	&	0.23	&	\(3000_{200}^{7000}\)	&	0.36	\\
0804+499	&	0804+499	&	626	&	FSRQ	&	\(1.5_{1.3}^{2.0}\)	&	\(6500_{1000}^{7000}\)	&	0.74	&	\(1.4_{1.3}^{1.6}\)	&	0.66	&	\(6500_{100}^{7000}\)	&	0.78	\\
0805-077	&	0805-077	&	154	&	FSRQ	&	\(2.7_{1.9}^{3.5}\)	&	\(4500_{800}^{7000}\)	&	0.89	&	\(2.2_{1.7}^{3.4}\)	&	0.75	&	\(4000_{500}^{7000}\)	&	0.86	\\
0814+425	&	0814+425	&	332	&	BLO	&	\(1.6_{1.3}^{3.1}\)	&	\(4500_{400}^{7000}\)	&	0.46	&	\(1.4_{1.2}^{1.5}\)	&	0.25	&	\(2000_{100}^{7000}\)	&	0.35	\\
0823+033	&	0823+033	&	146	&	BLO	&	\(2.9_{1.6}^{3.5}\)	&	\(1500_{400}^{7000}\)	&	0.93	&	\(1.7_{1.4}^{2.2}\)	&	0.52	&	\(500_{200}^{2500}\)	&	0.91	\\
0827+243	&	OJ248	&	671	&	FSRQ	&	\(2.2_{1.6}^{3.5}\)	&	\(2500_{500}^{7000}\)	&	0.69	&	\(1.7_{1.5}^{2.2}\)	&	0.45	&	\(1500_{400}^{7000}\)	&	0.60	\\
0829+046	&	0829+046	&	186	&	BLO	&	\(2.3_{1.8}^{3.5}\)	&	\(6000_{800}^{7000}\)	&	0.82	&	\(2.0_{1.6}^{3.5}\)	&	0.89	&	\(5500_{500}^{7000}\)	&	0.87	\\
0836+710	&	0836+710	&	1291	&	FSRQ	&	\(1.3_{1.2}^{1.4}\)	&	\(7000_{2000}^{7000}\)	&	0.48	&	\(1.2_{1.1}^{1.4}\)	&	0.61	&	\(6500_{100}^{7000}\)	&	0.59	\\
0851+202	&	OJ287	&	2938	&	BLO	&	\(2.0_{1.7}^{2.5}\)	&	\(3000_{800}^{7000}\)	&	0.61	&	\(1.6_{1.5}^{1.8}\)	&	0.36	&	\(2000_{800}^{7000}\)	&	0.61	\\
0859+470	&	0859+470	&	146	&	FSRQ	&	\(1.2_{1.0}^{3.5}\)	&	\(5000_{100}^{7000}\)	&	0.76	&	\(1.1_{1.0}^{1.7}\)	&	0.82	&	\(5000_{100}^{7000}\)	&	0.84	\\
0906+430	&	0906+430	&	269	&	GAL	&	\(1.0_{1.0}^{1.5}\)	&	\(7000_{100}^{7000}\)	&	0.22	&	\(1.0_{1.0}^{1.1}\)	&	0.22	&	\(800_{100}^{7000}\)	&	0.24	\\
0917+449	&	0917+449	&	155	&	FSRQ	&	\(1.7_{1.5}^{2.3}\)	&	\(6500_{1500}^{7000}\)	&	0.50	&	\(1.6_{1.4}^{1.8}\)	&	0.50	&	\(6500_{400}^{7000}\)	&	0.50	\\
0923+392	&	4C39.25	&	1419	&	FSRQ	&	&	&	&\( 1.6_{1.4}^{1.8}\)	&	0.17	&	&	\\
0945+408	&	0945+408	&	277	&	FSRQ	&	\(1.6_{1.5}^{2.0}\)	&	\(6500_{3000}^{7000}\)	&	0.25	&	\(1.5_{1.3}^{2.5}\)	&	0.45	&	\(7000_{500}^{7000}\)	&	0.34	\\
0953+254	&	0953+254	&	362	&	FSRQ	&	\(1.9_{1.5}^{3.5}\)	&	\(3000_{400}^{7000}\)	&	0.78	&	\(1.5_{1.4}^{1.8}\)	&	0.57	&	\(1500_{200}^{7000}\)	&	0.80	\\
0954+556	&	S40954+556	&	237	&	FSRQ	&	\(1.3_{1.0}^{3.5}\)	&	\(3500_{100}^{7000}\)	&	0.87	&	\(1.2_{1.0}^{1.8}\)	&	0.88	&	\(2500_{100}^{7000}\)	&	0.90	\\
0954+65	&	S40954+65	&	606	&	BLO	&	\(1.4_{1.2}^{1.8}\)	&	\(6000_{800}^{7000}\)	&	0.51	&	\(1.3_{1.1}^{1.7}\)	&	0.55	&	\(5500_{100}^{7000}\)	&	0.59	\\
1036+054	&	1036+054	&	144	&	FSRQ	&	\(3.1_{3.1}^{3.4}\)	&	\(1000_{800}^{1000}\)	&	0.10	&	&		&	&	\\
1049+215	&	1049+215	&	226	&	FSRQ	&	\(3.5_{2.0}^{3.5}\)	&	\(7000_{2000}^{7000}\)	&	0.83	&	\(3.5_{1.6}^{3.5}\)	&	0.96	&	\(7000_{800}^{7000}\)	&	0.84	\\
1055+018	&	1055+018	&	968	&	BLO	&	\(1.8_{1.7}^{2.4}\)	&	\(7000_{1500}^{7000}\)	&	0.89	&	\(1.6_{1.5}^{1.8}\)	&	0.70	&	\(7000_{800}^{7000}\)	&	0.87	\\
1101+384	&	MARK421	&	1511	&	BLO	&	\(1.6_{1.3}^{3.5}\)	&	\(6500_{1000}^{7000}\)	&	0.52	&	\(1.5_{1.2}^{3.1}\)	&	0.76	&	\(5500_{100}^{7000}\)	&	0.69	\\
1150+497	&	1150+497	&	212	&	FSRQ	&	\(1.8_{1.4}^{3.5}\)	&	\(6500_{400}^{7000}\)	&	0.76	&	\(1.6_{1.3}^{3.5}\)	&	0.73	&	\(4000_{200}^{7000}\)	&	0.74	\\
1156+295	&	4C29.45	&	1887	&	FSRQ	&	\(2.5_{1.8}^{3.5}\)	&	\(1500_{500}^{7000}\)	&	0.76	&	\(1.7_{1.5}^{2.0}\)	&	0.35	&	\(1000_{400}^{4000}\)	&	0.54	\\
1219+285	&	ON231	&	566	&	BLO	&	\(2.7_{1.8}^{3.5}\)	&	\(6500_{1500}^{7000}\)	&	0.85	&	\(1.9_{1.5}^{3.5}\)	&	0.97	&	\(7000_{500}^{7000}\)	&	0.92	\\
1222+216	&	PKS1222+216	&	931	&	FSRQ	&	\(1.7_{1.6}^{2.1}\)	&	\(7000_{2500}^{7000}\)	&	0.46	&	\(1.6_{1.4}^{1.9}\)	&	0.66	&	\(7000_{800}^{7000}\)	&	0.53	\\
1226+023	&	3C273	&	2267	&	FSRQ	&	\(2.3_{2.1}^{2.9}\)	&	\(7000_{2000}^{7000}\)	&	0.62	&	\(1.9_{1.8}^{2.1}\)	&	0.43	&	\(4000_{2000}^{7000}\)	&	0.63	\\
1253-055	&	3C279	&	2293	&	FSRQ	&	\(1.8_{1.7}^{2.1}\)	&	\(4500_{2000}^{7000}\)	&	0.63	&	\(1.6_{1.5}^{1.7}\)	&	0.45	&	\(3000_{800}^{7000}\)	&	0.66	\\
1308+326	&	1308+326	&	918	&	BLO	&	\(1.9_{1.8}^{2.2}\)	&	\(6500_{3500}^{7000}\)	&	0.30	&	\(1.7_{1.5}^{1.9}\)	&	0.47	&	\(5000_{1500}^{7000}\)	&	0.38	\\
1324+224	&	1324+224	&	322	&	FSRQ	&	\(2.1_{1.4}^{3.5}\)	&	\(2000_{400}^{7000}\)	&	0.87	&	\(1.6_{1.4}^{1.9}\)	&	0.56	&	\(1500_{200}^{7000}\)	&	0.81	\\
1334-127	&	1334-127	&	445	&	FSRQ	&	\(1.8_{1.7}^{2.0}\)	&	\(7000_{3500}^{7000}\)	&	0.18	&	\(1.7_{1.6}^{1.8}\)	&	0.18	&	\(5500_{2500}^{7000}\)	&	0.18	\\
1406-076	&	PKS1406-076	&	261	&	FSRQ	&	\(3.5_{1.7}^{3.5}\)	&	\(7000_{1500}^{7000}\)	&	0.81	&	\(2.2_{1.5}^{3.5}\)	&	0.9	&	\(7000_{400}^{7000}\)	&	0.86	\\
1413+135	&	PKS1413+135	&	670	&	BLO	&	\(2.7_{2.1}^{3.5}\)	&	\(7000_{1500}^{7000}\)	&	0.73	&	\(2.1_{1.7}^{2.9}\)	&	0.93	&	\(5000_{1000}^{7000}\)	&	0.83	\\
1418+546	&	OQ530	&	503	&	BLO	&	\(1.8_{1.7}^{1.9}\)	&	\(6500_{6500}^{7000}\)	&	0.11	&	\(1.6_{1.5}^{1.8}\)	&	0.32	&	\(5000_{2500}^{7000}\)	&	0.25	\\
1502+106	&	PKS1502+106	&	815	&	FSRQ	&	\(2.2_{1.9}^{2.9}\)	&	\(6000_{2000}^{7000}\)	&	0.27	&	\(2.0_{1.6}^{3.5}\)	&	0.51	&	\(6000_{1000}^{7000}\)	&	0.36	\\
1510-089	&	PKS1510-089	&	1236	&	FSRQ	&	\(1.7_{1.5}^{2.9}\)	&	\(6000_{400}^{7000}\)	&	0.57	&	\(1.5_{1.4}^{1.8}\)	&	0.53	&	\(500_{200}^{3000}\)	&	0.81	\\
1538+149	&	4C14.60	&	402	&	BLO	&	\(1.6_{1.4}^{3.5}\)	&	\(7000_{800}^{7000}\)	&	0.91	&	\(1.4_{1.2}^{1.8}\)	&	0.90	&	\(6500_{100}^{7000}\)	&	0.92	\\
1546+027	&	1546+027	&	249	&	FSRQ	&	\(1.9_{1.6}^{3.5}\)	&	\(6500_{800}^{7000}\)	&	0.89	&	\(1.7_{1.4}^{3.1}\)	&	0.89	&	\(6000_{400}^{7000}\)	&	0.89	\\
1606+106	&	1606+106	&	632	&	FSRQ	&	\(2.1_{1.9}^{2.7}\)	&	\(7000_{2500}^{7000}\)	&	0.30	&	\(1.9_{1.6}^{2.4}\)	&	0.45	&	\(6500_{1500}^{7000}\)	&	0.43	\\
1611+343	&	DA406	&	998	&	FSRQ	&	\(1.9_{1.7}^{2.3}\)	&	\(7000_{2500}^{7000}\)	&	0.50	&	\(1.7_{1.5}^{2.0}\)	&	0.61	&	\(5500_{1500}^{7000}\)	&	0.58	\\
1633+382	&	4C38.41	&	1500	&	FSRQ	&	\(1.9_{1.7}^{2.1}\)	&	\(6000_{2500}^{7000}\)	&	0.48	&	\(1.6_{1.5}^{1.8}\)	&	0.48	&	\(5500_{1500}^{7000}\)	&	0.50	\\
1637+574	&	1637+574	&	772	&	FSRQ	&	\(1.6_{1.4}^{2.0}\)	&	\(6000_{1500}^{7000}\)	&	0.57	&	\(1.4_{1.3}^{1.7}\)	&	0.41	&	\(3000_{400}^{7000}\)	&	0.52	\\
1638+398	&	1638+398	&	269	&	FSRQ	&	\(1.6_{1.2}^{3.5}\)	&	\(4000_{200}^{7000}\)	&	0.69	&	\(1.4_{1.2}^{1.9}\)	&	0.61	&	\(6000_{100}^{7000}\)	&	0.66	\\
1641+399	&	3C345	&	2082	&	FSRQ	&	\(2.2_{2.2}^{2.3}\)	&	\(7000_{6500}^{7000}\)	&	0.12	&	\(1.9_{1.7}^{2.1}\)	&	0.41	&	\(7000_{3000}^{7000}\)	&	0.22	\\
1642+690	&	1642+690	&	255	&	FSRQ	&	\(1.9_{1.5}^{2.6}\)	&	\(7000_{1500}^{7000}\)	&	0.67	&	\(1.7_{1.4}^{2.6}\)	&	0.76	&	\(7000_{400}^{7000}\)	&	0.69	\\
1652+398	&	MARK501	&	1932	&	BLO	&	\(1.3_{1.2}^{3.3}\)	&	\(6500_{4000}^{7000}\)	&	0.22	&	\(1.3_{1.1}^{3.3}\)	&	0.51	&	\(4500_{100}^{7000}\)	&	0.35	\\
1725+044	&	PKS1725+044	&	276	&	FSRQ	&	\(1.7_{1.2}^{3.5}\)	&	\(1500_{200}^{7000}\)	&	0.90	&	\(1.3_{1.1}^{1.5}\)	&	0.57	&	\(200_{100}^{4000}\)	&	0.72	\\
1730-130	&	1730-130	&	565	&	FSRQ	&	\(2.1_{1.8}^{2.8}\)	&	\(7000_{2000}^{7000}\)	&	0.61	&	\(1.9_{1.6}^{2.4}\)	&	0.76	&	\(7000_{1500}^{7000}\)	&	0.70	\\
1739+52	&	S41739+52	&	346	&	FSRQ	&	\(1.4_{1.4}^{1.5}\)	&	\(5500_{5000}^{7000}\)	&	0.13	&	\(1.3_{1.2}^{1.4}\)	&	0.19	&	\(4500_{800}^{7000}\)	&	0.18	\\
1741-038	&	1741-038	&	1073	&	FSRQ	&	\(1.9_{1.7}^{2.2}\)	&	\(6000_{2000}^{7000}\)	&	0.63	&	\(1.6_{1.5}^{1.9}\)	&	0.63	&	\(7000_{2000}^{7000}\)	&	0.69	\\
1749+096	&	PKS1749+096	&	1270	&	BLO	&	\(2.1_{1.6}^{2.8}\)	&	\(1000_{400}^{7000}\)	&	0.90	&	\(1.5_{1.4}^{1.5}\)	&	0.19	&	\(400_{200}^{800}\)	&	0.58	\\
1803+784	&	S51803+784	&	363	&	BLO	&	\(1.1_{1.0}^{1.6}\)	&	\(7000_{400}^{7000}\)	&	0.78	&	\(1.0_{1.0}^{1.2}\)	&	0.86	&	\(5000_{100}^{7000}\)	&	0.90	\\
1807+698	&	3C371.0	&	438	&	BLO	&	\(1.2_{1.1}^{1.4}\)	&	\(6000_{1000}^{7000}\)	&	0.36	&	\(1.1_{1.0}^{1.3}\)	&	0.44	&	\(2500_{100}^{7000}\)	&	0.48	\\
1823+568	&	4C56.27	&	259	&	BLO	&	\(1.5_{1.2}^{1.9}\)	&	\(6500_{1500}^{7000}\)	&	0.70	&	\(1.4_{1.2}^{1.9}\)	&	0.81	&	\(4000_{100}^{7000}\)	&	0.76	\\
1828+487	&	1828+487	&	360	&	FSRQ	&	\(1.7_{1.5}^{3.5}\)	&	\(6000_{800}^{7000}\)	&	0.49	&	\(1.5_{1.4}^{1.9}\)	&	0.49	&	\(5500_{400}^{7000}\)	&	0.48	\\
1845+797	&	3C390.3	&	241	&	GAL	&	\(1.0_{1.0}^{1.9}\)	&	\(3500_{100}^{7000}\)	&	0.25	&	\(1.0_{1.0}^{1.0}\)	&	0.12	&	\(6000_{100}^{7000}\)	&	0.24	\\
1901+319	&	1901+319	&	365	&	FSRQ	&	\(1.5_{1.2}^{2.1}\)	&	\(6500_{1500}^{7000}\)	&	0.43	&	\(1.4_{1.2}^{3.5}\)	&	0.73	&	\(5500_{100}^{7000}\)	&	0.62	\\
1928+738	&	1928+738	&	413	&	FSRQ	&		&	&		&		&	&		&\\
1954+513	&	1954+513	&	446	&	FSRQ	&	\(1.3_{1.1}^{1.6}\)	&	\(3500_{800}^{7000}\)	&	0.28	&	\(1.2_{1.1}^{1.4}\)	&	0.25	&	\(6500_{100}^{7000}\)	&	0.24	\\
2005+403	&	2005+403	&	1104	&	FSRQ	&		&		&		&	\(1.6_{1.5}^{2.0}\)	&	0.31	&	&		\\
2007+77	&	S52007+77	&	297	&	FSRQ	&	\(1.4_{1.1}^{3.5}\)	&	\(1000_{100}^{7000}\)	&	0.28	&	\(1.1_{1.1}^{1.2}\)	&	0.15	&	\(100_{100}^{7000}\)	&	0.22	\\
2021+614	&	2021+614	&	394	&	GAL	&	\(1.0_{1.0}^{2.7}\)	&	\(800_{100}^{7000}\)	&	0.90	&	\(1.0_{1.0}^{1.1}\)	&	0.51	&	\(200_{100}^{7000}\)	&	0.71	\\
2022+171	&	2022+171	&	363	&	FSRQ	&		&	&	&		&		&	&		\\
2022-077	&	PKS2022-077	&	136	&	BLO/FSRQ	&	\(1.6_{1.3}^{3.5}\)	&	\(5000_{200}^{7000}\)	&	0.63	&	\(1.5_{1.3}^{1.8}\)	&	0.49	&	\(3500_{100}^{7000}\)	&	0.61	\\
2037+511	&	2037+511	&	351	&	FSRQ	&	\(1.5_{1.4}^{1.7}\)	&	\(7000_{3500}^{7000}\)	&	0.23	&	\(1.5_{1.3}^{1.6}\)	&	0.46	&	\(7000_{800}^{7000}\)	&	0.33	\\
2121+053	&	2121+053	&	252	&	FSRQ	&	\(1.6_{1.4}^{2.3}\)	&	\(7000_{1500}^{7000}\)	&	0.63	&	\(1.5_{1.3}^{1.9}\)	&	0.80	&	\(6500_{400}^{7000}\)	&	0.69	\\
2131-021	&	2131-021	&	213	&	BLO/FSRQ	&	\(1.6_{1.5}^{1.8}\)	&	\(6000_{3500}^{7000}\)	&	0.17	&	\(1.4_{1.3}^{1.7}\)	&	0.32	&	\(7000_{500}^{7000}\)	&	0.25	\\
2134+004	&	2134+004	&	763	&	FSRQ	&	\(1.4_{1.3}^{1.9}\)	&	\(6500_{800}^{7000}\)	&	0.76	&	\(1.3_{1.2}^{1.4}\)	&	0.64	&	\(7000_{500}^{7000}\)	&	0.25	\\
2136+141	&	2136+141	&	358	&	FSRQ	&	\(1.7_{1.3}^{3.5}\)	&	\(3500_{400}^{7000}\)	&	0.97	&	\(1.4_{1.2}^{2.2}\)	&	0.89	&	\(3500_{100}^{7000}\)	&	0.96	\\
2144+092	&	2144+092	&	191	&	FSRQ	&	\(2.6_{1.4}^{3.5}\)	&	\(800_{400}^{7000}\)	&	0.92	&	\(1.5_{1.3}^{1.8}\)	&	0.34	&	\(200_{100}^{800}\)	&	0.72	\\
2145+067	&	2145+067	&	1079	&	FSRQ	&		&		&	&	\(1.8_{1.7}^{1.9}\)	&	0.33	&	&	\\
2200+420	&	BL Lac	&	3639	&	BLO	&	\(1.9_{1.7}^{2.2}\)	&	\(7000_{2000}^{7000}\)	&	0.62	&	\(1.7_{1.5}^{2.0}\)	&		0.70	&	\(4500_{1000}^{7000}\)	&	0.73	\\
2201+171	&	2201+171	&	214	&	BLO/FSRQ	&	\(3.4_{2.0}^{3.5}\)	&	\(1500_{800}^{7000}\)	&	0.41	&	\(2.1_{1.8}^{2.2}\)	&	0.16	&	\(800_{500}^{3500}\)	&	0.34	\\
2201+315	&	2201+315	&	1092	&	FSRQ	&	\(1.9_{1.7}^{2.5}\)	&	\(5000_{1500}^{7000}\)	&	0.52	&	\(1.6_{1.5}^{1.9}\)	&	0.52	&	\(3500_{800}^{7000}\)	&	0.56	\\
2216-038	&	2216-038	&	150	&	FSRQ	&	\(2.5_{1.6}^{3.5}\)	&	\(3500_{800}^{7000}\)	&	0.64	&	\(1.6_{1.4}^{2.1}\)	&	0.41	&	\(800_{200}^{4000}\)	&	0.49	\\
2223-052	&	3C446	&	754	&	BLO	&	\(2.1_{2.0}^{2.3}\)	&	\(7000_{4500}^{7000}\)	&	0.24	&	\(1.8_{1.7}^{2.1}\)	&	0.62	&	\(6500_{2500}^{7000}\)	&	0.41	\\
2227-088	&	2227-088	&	298	&	FSRQ	&	\(2.1_{1.5}^{3.5}\)	&	\(2500_{400}^{7000}\)	&	0.97	&	\(1.7_{1.4}^{2.2}\)	&	0.68	&	\(800_{200}^{7000}\)	&	0.96	\\
2230+114	&	2230+114	&	1202	&	FSRQ	&	\(2.3_{1.7}^{2.6}\)	&	\(1000_{800}^{7000}\)	&	0.33	&		&		&	\(500_{400}^{1000}\)	&	0.20	\\
2234+282	&	2234+282	&	313	&	FSRQ	&	\(1.4_{1.1}^{3.5}\)	&	\(3000_{100}^{7000}\)	&	0.99	&	\(1.2_{1.0}^{1.6}\)	&	0.99	&	\(2500_{100}^{7000}\)	&	0.99	\\
2251+158	&	3C454.3	&	3154	&	FSRQ	&	\(2.3_{2.0}^{2.8}\)	&	\(3000_{1000}^{7000}\)	&	0.79	&	\(1.8_{1.6}^{2.0}\)	&	0.32	&	\(1000_{500}^{3000}\)	&	0.75	\\
2344+092	&	2344+092	&	172	&	GAL/FSRQ	&	\(3.2_{1.9}^{3.5}\)	&	\(7000_{1500}^{7000}\)	&	0.88	&	\(2.2_{1.6}^{3.5}\)	&	0.91	&	\(6000_{800}^{7000}\)	&	0.89	\\
2351+456	&	4C45.51	&	265	&	FSRQ	&	\(2.0_{1.6}^{3.5}\)	&	\(7000_{800}^{7000}\)	&	0.39	&	\(1.8_{1.5}^{2.4}\)	&	0.47	&	\(7000_{400}^{7000}\)	&	0.43	\\
\hline
\hline
\end{longtable}
\tablefoot{For each source we give the total number of data points (N) in the light curve, the best-fit bending power-law slope \(\beta_{bpl}\) and bend timescale \(x_b\)\footnote{Only the timescales for the 11 constrained sources should be considered reliable. For the other sources more observations are needed.} using Eq. \ref{bpl}, as well as the corresponding p-values (\(p_{bpl}\)). For the best-fit simple power-law parameters we give both the slope \(\beta_{spl}\) and the best-fit p-values (\(p_{spl}\)). For the case with \(\beta_{low}\) = 1 using Eq. \ref{hfbpl}, we give the best-fit bending power-law bend timescale (\(x_{b1}\)) where \(\beta_{high}\) = \(\beta_{bpl, best}\). The corresponding p-values are given by \(p_{bpl_1}\). All values are reported with their 90 \% confidence regions, where the slope and timescale constraints for the bending power-law fits (Eq. 4) are the minimum and maximum values obtained from some combination of the two parameters within the bending power-law fit 90 \% confidence region. A missing value in the table indicates a rejection confidence of over 90 \%. The probed limits for the timescales were \(x_{b,min}\) = 100 days and \(x_{b,max}\) = 7000 days, and the probed limits for the slopes were \(\beta_{min}\) = 1 and \(\beta_{max}\) = 3.5.}

\end{appendix}

\end{document}